\documentclass{article}
\usepackage{bm,latexsym,amsmath,amsfonts,amssymb,fancyhdr,color,graphicx,slashed,cite}
\usepackage[a4paper,bottom=2.5cm,top=3.cm,head=5mm,width=17cm,dvipdfm]{geometry}
\usepackage[usenames,dvipsnames,svgnames,table]{xcolor}
\usepackage[colorlinks=true,
            linkcolor=blue,
            urlcolor=blue,
            citecolor=green,          
            bookmarks=true,
            bookmarksnumbered=true,
            breaklinks=true,
            pdfpagemode=Fullscreen,
            pdfstartview=FitBH]{hyperref}

\usepackage[dotinlabels]{titletoc}
 \usepackage[normalem]{ulem}
\usepackage{float}
\usepackage{extarrows}
\usepackage{xcolor}
\usepackage{cancel}
\usepackage[bbgreekl]{mathbbol}
\usepackage{bbm}
\usepackage{siunitx}
\usepackage{enumitem}
\usepackage{tikz,pgfplots}
\usetikzlibrary{positioning}
\usepackage[title]{appendix}

\usepackage[capitalise]{cleveref}
\usepackage[version=4]{mhchem}
\usepackage[font=small]{caption}

\usepackage{multirow}
\usepackage{feynmf}
\numberwithin{equation}{section}
\allowdisplaybreaks[4]

\def\lsim{\mathrel{\raise.3ex\hbox{$<$\kern-.75em\lower1ex\hbox{$\sim$}}}}
\def\gsim{\mathrel{\raise.3ex\hbox{$>$\kern-.75em\lower1ex\hbox{$\sim$}}}}

\usepackage{tikz} 
\usepackage{tkz-euclide}
\usetikzlibrary{backgrounds} 
\usetikzlibrary{decorations.pathmorphing}
\usetikzlibrary{arrows.meta}
\tikzset{
mystyle/.style={line width=1, baseline, scale=0.6, every node/.style={scale=1}},
v/.style={decorate, draw, decoration={snake, segment length=2.mm, amplitude=0.5mm}},
f/.style={draw, decoration={markings,mark=at position #1 with {\arrow[]{Latex[length=1.5mm,width=1.5mm]}}},
    postaction={decorate},node contents=#1},
f/.default=.6,
fb/.style={draw,decoration={markings,mark=at position #1 with {\arrowreversed[]{Latex[length=1.5mm,width=1.5mm]}}},
    postaction={decorate},node contents=#1},
fb/.default=.6,
s/.style={dashed,draw, decoration={markings,mark=at position #1 with {\arrow[]{Latex[length=1.5mm,width=1.5mm]}}},
    postaction={decorate},node contents=#1},
s/.default=.6,    
sb/.style={dashed,draw,decoration={markings,mark=at position #1 with {\arrowreversed[]{Latex[length=1.5mm,width=1.5mm]}}},
    postaction={decorate},node contents=#1},
sb/.default=.4,
snar/.style={dashed,draw,line width =1.25pt},
cross/.style={cross out, draw=black, minimum size=2*(#1-\pgflinewidth), inner sep=0pt, outer sep=0pt}, 
         }

\newcommand{\tr}{\mbox{Tr}}
\newcommand{\calL}{ {\cal L} }
\newcommand{\calO}{ {\cal O} }
\newcommand{\chpt}{\chi{\rm PT}}
\newcommand{\NR}{{\tt NR}}
\newcommand{\ER}{{\tt ER}}
\newcommand{\OTa}{\mathcal{O}_{\chi q}^{\tt T1}}
\newcommand{\OTb}{\mathcal{O}_{\chi q}^{\tt T2}}

\usepackage[titles]{tocloft}

\providecommand{\keywords}[1]
{
  \small	
  \textbf{KEYWORDS:} #1
}

\begin{document} 
\fontsize{10pt}{12pt}\selectfont

\title{\textbf{Comprehensive constraints on fermionic dark matter-quark tensor interactions in direct detection experiments}}

\author{
{Jin-Han Liang~$^{a,b}$}\footnote{jinhanliang@m.scnu.edu.cn},\,
{Yi Liao~$^{a,b}$}\footnote{liaoy@m.scnu.edu.cn},\,
{Xiao-Dong Ma~$^{a,b}$}\footnote{maxid@scnu.edu.cn}\, and 
{Hao-Lin Wang~$^{a,b}$}\footnote{whaolin@m.scnu.edu.cn}
 \\[3mm]
{\small $^a$~Key Laboratory of Atomic and Subatomic Structure and Quantum Control (MOE),}\\
{\small Guangdong Basic Research Center of Excellence for Structure and Fundamental Interactions of Matter, } \\
{\small Institute of Quantum Matter, South China Normal University, Guangzhou 510006, China}  \\[1mm]
{\small $^b$~Guangdong-Hong Kong Joint Laboratory of Quantum Matter,}\\
{\small Guangdong Provincial Key Laboratory of Nuclear Science, }\\
{\small Southern Nuclear Science Computing Center, South China Normal University, Guangzhou 510006, China }
}

\date{}
\maketitle

\vspace{-0.75cm}

%
%%%%%%%%%%%%%%%%%
\begin{abstract}
%%%%%%%%%%%%%%%%%
Effective field theory (EFT) provides a model-independent framework for interpreting the results of dark matter (DM) direct detection experiments. 
In this study, we demonstrate that the two fermionic DM-quark tensor operators, 
$(\bar{\chi} i\sigma^{\mu\nu} \gamma^5 \chi) (\bar{q} \sigma_{\mu\nu}q)$ and $(\bar{\chi} \sigma^{\mu\nu} \chi) (\bar{q} \sigma_{\mu\nu} q)$, 
can contribute to the DM electric and magnetic dipole moments via nonperturbative QCD effects, 
in addition to the well-studied contact DM-nucleon operators. 
We then investigate the constraints on these two operators by considering both the contact
and the dipole contributions using the XENON1T nuclear recoil and Migdal effect data.
We also recast other existing bounds on the DM dipole operators, 
derived from electron and nuclear recoil measurements in various direct detection experiments,
as constraints on the two tensor operators. For $m_\chi \lesssim 1\,\si{GeV}$,
our results significantly extend the reach of constraints on the DM-quark tensor operators
to masses as low as $5\,\si{MeV}$, with the bound exceeding that obtained by the Migdal effect
with only contact interactions by an order of magnitude or so. 
In particular, for the operator $(\bar{\chi} \sigma^{\mu\nu}i\gamma_5 \chi) (\bar{q} \sigma_{\mu\nu}q)$ with DM mass $m_\chi \gtrsim 10\,\si{GeV}$, the latest PandaX constraint on the DM electric dipole moment puts more stringent bounds than the previous direct detection limit. We also discuss briefly the constraints coming from experiments other than direct detection.
\end{abstract}
%%%%%%%%%%%%%%%

%

\newpage
\hypersetup{linkcolor=black}
\tableofcontents
\hypersetup{linkcolor=red}

%%%%%%%%%%%%%%%%%%%%%%%%%%%%%%
\section{Introduction}
%%%%%%%%%%%%%%%%%%%%%%%%%%%%%%

Although dark matter (DM) makes up about a quarter of the total energy density of the Universe, 
its particle properties still remain unknown today \cite{Bertone:2004pz,Feng:2010gw}. 
One of the theoretically motivated candidates is the weakly interacting massive particle (WIMP),
which can meet the required properties to explain the DM conundrum, 
and at the same time, have a detectable possibility.
During the past two decades, although a great amount of 
theoretical and experimental efforts have been dedicated to searches for the WIMPs,
the DM direct detection (DMDD) experiments have not found any positive signal 
but constrained the DM-nucleus cross section to an unprecedented level \cite{Roszkowski:2017nbc,Schumann:2019eaa}.
However, due to the kinematic restriction of DM-nucleus elastic scattering, 
the DM-nucleon interaction remains less constrained below
the ${\cal O}(1\,\rm GeV)$ DM mass region from direct detection experiments
using nuclear recoil ({\tt NR}) signals. 
To address this limitation, inelastic processes are taken into account, 
for instance, bremsstrahlung processes \cite{Kouvaris:2016afs} and the Migdal effect \cite{Migdal:1941,Ibe:2017yqa}. 
Nevertheless, even with the improvement from inelastic processes, 
the constraints are still limited to a mass up to 40 MeV or so \cite{LUX:2018akb,XENON:2019zpr,EDELWEISS:2019vjv,Tomar:2022ofh,DarkSide:2022dhx,SuperCDMS:2023sql}. 
For lighter DM particles at the MeV scale, meaningful constraints require considerations like boosted DM scenarios \cite{Cappiello:2018hsu,Bringmann:2018cvk} or 
novel low-threshold detectors, see the review paper \cite{Kahn:2021ttr} and references therein.
In contrast to $\NR$ experiments searching for the DM-nucleon interaction, 
the DM-electron interaction offers a more powerful alternative of probing low-mass DM particles
through the electron recoil ($\ER$) signal \cite{Essig:2011nj,Essig:2012yx}.
This is due to the significantly smaller mass of the electron as compared to a typical nucleus, 
allowing it to gain easily the recoil energy from a light DM particle. 
For instance, the single-electron search conducted by XENON1T has the capability of
exploring a DM mass as low as approximately 5 MeV \cite{XENON:2021qze}.

Due to the small momentum transfer (less than a few MeV)
for DMDD experiments,
it is preferable to adopt the low energy effective field theory (LEFT) approach
to study the interactions between DM and standard model particles,
which basically does not rely on the details of ultraviolet (UV) models \cite{Kang:2018odb,Kang:2018rad,Catena:2019gfa}. 
The starting point for DM EFT is the DM-quark, -lepton, or -photon/gluon interactions
at leading order that are color and electric charge neutral. 
In the Dirac DM case, the leading operators appear at dimension 5 and 6,
and take the general form, $\bar \chi \sigma_{\mu\nu}(i\gamma_5) \chi F^{\mu\nu}$
and $(\bar \chi \Gamma \chi)(\bar \psi \Gamma' \psi)$,
where $\chi$ represents the fermionic Dirac-type DM,
$\psi$ the quarks or leptons,
and $F^{\mu\nu}$ the electromagnetic field strength tensor.
For the interest of direct detection, 
$\psi$ is usually taken to be the up, down, strange quark, and the electron.
For the $\NR$, the DM-nucleon interaction naturally arises from DM-quark and DM-gluon operators
through nonperturbative matching via chiral perturbation theory ($\chi$PT) \cite{Bishara:2017pfq}.

In this work, we explore the DM-photon interactions induced by nonperturbative QCD effects from the DM-quark interactions. 
This will provide new ways to constrain the DM-quark interactions. Similar ideas have already been used
in the study of flavor-violated radiative decays of charged leptons
and neutrino electromagnetic (EM) moments \cite{Dekens:2018pbu, Chen:2022xkk}.
In particular, we consider the two tensor operators,
$(\bar{\chi} i\sigma^{\mu\nu} \gamma^5 \chi) (\bar{q} \sigma_{\mu\nu}q)$ and $(\bar{\chi} \sigma^{\mu\nu} \chi)( \bar{q} \sigma_{\mu\nu} q)$, 
which not only induce the short-distance (SD) DM-nucleon operators covered in most direct detection studies, 
but also generate the DM electric and magnetic dipole moment operators, 
$(\bar{\chi} i \sigma_{\mu\nu} \gamma_5 \chi) F^{\mu\nu}$ (edm) and $(\bar{\chi} \sigma_{\mu\nu} \chi) F^{\mu\nu}$ (mdm), 
which also contribute to the direct detection via the long-distance (LD) photon mediator. 

In previous calculations for DMDD constraints on the above two tensor operators, 
only the DM-nucleus scattering induced by the SD operators is considered \cite{Kang:2018odb,Kang:2018rad,Tomar:2022ofh}. 
In this work, we utilize the XENON1T experiment as a benchmark experiment to investigate comprehensively the constraints from both SD and LD contributions. 
We find that there are interesting interference effects between the two in the DM-nucleus scattering and the Migdal effect. 
Due to the induced dipole interactions, we also investigate the constraints from DM-electron scattering. 
Remarkably, owing to the excellent sensitivity to low-mass DM, 
the constraints from $\ER$ via this LD effect significantly extend to low-mass (from GeV to MeV) DM. 
In addition, we collect other existing direct detection and non-direct-detection constraints on the DM dipole operators and recast them into constraints on the DM-quark tensor operators. 
In our analysis, we will consider the cases of whether the flavor SU(3) symmetry is imposed or not. 
For the flavor conserving case, a universal Wilson coefficient is assumed for the operators involving the $u,d,s$ quarks 
(here the corresponding quark mass is attached to the operator as usually practiced in the literature \cite{Bishara:2017pfq}); 
and for the non-conserving case, the contributions from the three quarks are considered separately.

The paper is organized as follows. \cref{sec:match} is dedicated to nonperturbative chiral matching of the two DM-quark tensor operators onto the DM-photon and DM-nucleon interactions. 
We then discuss in \cref{sec:xenon1} the constraints on these operators from the $\NR$, the Migdal effect, and the $\ER$, based on the XENON1T data. 
The full constraints and comparisons with the literature 
for direct detection experiments
are given in \cref{sec:full_cons}.  
 In \cref{sec:nonDD},
we further discuss constraints from non-direct-detection 
experiments and give an example of UV completion for the
two DM-quark tensor operators.
Our concluding remarks are presented in \cref{sec:conc}. 
The relevant nuclear form factors are collected in \cref{app:sform}.

%%%%%%%%%%%%%%%%%%%%%%%%%%%%%%
\section{Nonperturbative matching of DM-quark interactions}
\label{sec:match}
%%%%%%%%%%%%%%%%%%%%%%%%%%%%%%

Since the transferred momentum in DMDD experiments is limited to a few hundreds of MeV,
we can generically describe the interactions between DM and standard model (SM) light fields within the framework of LEFT.
For DM direct detection, the complete set of operators with fermion and scalar DM particles
up to canonical mass dimension 7 have been classified in \cite{Bishara:2017pfq,Brod:2017bsw,Goodman:2010ku,Li:2021phq, Liang:2023yta}.
Here we are particularly interested in the two tensor operators for the Dirac fermionic DM
which have not got much attention except for a few studies focusing on $\NR$ signals
and the Migdal effect induced by the SD DM-nucleon interactions \cite{Kang:2018rad,Tomar:2022ofh}. Following the convention in \cite{Bishara:2017pfq}, they are parameterized by,
\begin{align}
\label{eq:op}
\OTa  = 
m_q\left(\bar{\chi} \sigma^{\mu \nu} \chi\right)\left(\bar{q} \sigma_{\mu \nu} q\right), \quad
\OTb  =  
m_q\left(\bar{\chi} i \sigma^{\mu \nu} \gamma_5 \chi\right)\left(\bar{q} \sigma_{\mu \nu} q\right),
\end{align}  
where $q$ represents the three light quarks $u,d,s$ of mass $m_q$ relevant to the direct detection.
For each operator, there is a corresponding unknown Wilson coefficient whose magnitude is parameterized as 
$|C_{\chi q}^{\tt T1(T2)}| \equiv 1/ \Lambda^3$ where $\Lambda$ is an effective scale related to some unknown UV physics. 
In the following, we demonstrate that these operators not only contribute to 
DM-nucleon local interactions but also the DM magnetic 
and electric dipole moment operators, 
$(\bar{\chi} \sigma_{\mu\nu} \chi) F^{\mu\nu}$ and 
$(\bar{\chi} i \sigma_{\mu\nu} \gamma_5 \chi) F^{\mu\nu}$, 
through nonperturbative QCD effects. 
These nonpurterbative dipole contributions to DM direct detection will help significantly extend the sensitivity to low mass DM. And they can be systematically extracted through matching 
within the framework of the (baryon) chiral perturbation theory ((B)$\chpt$) of QCD at low energy. 
For applications of (B)$\chpt$ in the description of DM direct detection,
see, for instance, Refs.\,\cite{Bishara:2016hek,Bishara:2017pfq}.
Other DM-quark operators involving (pseudo-)scalar or (axial-)vector currents, such as the commonly discussed
$\bar{\chi}\chi \bar{q} (\gamma_5) q$ and $\bar{\chi}\gamma^\mu \chi \bar{q} \gamma_\mu (\gamma_5) q$,
do not exhibit this unique nonperturbative matching to DM dipole moments but rather would generate operators with at least two photon fields due to QED gauge and parity symmetries. Their contributions to DM direct detection can be safely ignored due to suppression from loop factor and additional QED couplings.

Our starting point is the quark level Lagrangian with external sources
\begin{align}
\label{eq:exter}
\calL = \mathcal{L}_{\mathrm{QCD}}
+\overline{q_L} l_\mu \gamma^\mu q_L+\overline{q_R} r_\mu \gamma^\mu q_R
-\left[\overline{q_R}(s+i p) q_L-\overline{q_R} t^{\mu \nu} \sigma_{\mu \nu} q_L
+\text { h.c. }\right],
\end{align}
where $\mathcal{L}_{\rm QCD}$ is the QCD Lagrangian for the $u,~d,~s$ quarks in the chiral limit.
The external sources, $l_\mu,\,r_\mu,\,s,\,p$, and $t^{\mu\nu}$, are $3\times 3$ matrices in flavor space,
which contain non-strongly interacting fields like the leptons, photon,
and DM that interact with quarks. For the two DM-quark interactions
with a tensor quark current in \cref{eq:op}, the corresponding tensor external source is given by
\begin{align}
t^{\mu \nu} = P_L^{\mu\nu\alpha\beta} \bar{t}_{\alpha\beta},
\end{align}
where the first factor on the right-hand side is a tensor chiral projection operator
\cite{Cata:2007ns} defined as
\begin{align}
P_{L}^{\mu \nu \alpha \beta} = 
\frac{1}{4}\left(g^{\mu \alpha} g^{\nu \beta}-g^{\mu \beta} g^{\nu \alpha} 
- i \epsilon^{\mu \nu \alpha \beta}\right),
\end{align}
and $\bar t^{\mu\nu}$ is related to the DM tensor currents 
and couplings $C_{\chi q}^{\tt T1(T2)}$, which is a diagonal matrix in flavor space in our consideration, 
\begin{align}
(\bar t^{\mu\nu})_{qq} =
\mathcal{C}^{\tt T1}_{\chi q} m_q (\bar \chi \sigma^{\mu\nu} \chi)
 + \mathcal{C}^{\tt T2}_{\chi q} m_q (\bar \chi i \sigma^{\mu\nu}\gamma_5 \chi).   
\end{align}

The building blocks of chiral matching of the Lagrangian in \cref{eq:exter} 
consist of the pseudo Nambu-Goldstone boson (pNGB) matrix $U$, 
the baryon octet fields $B$, and external sources. 
We begin with the pure mesonic chiral Lagrangian that will lead to DM EM moments directly.
The leading order Lagrangian appears at $\calO(p^2)$ 
in chiral power counting \cite{Gasser:1983yg,Gasser:1984gg},
\begin{align}
\label{p2l}
\mathcal{L}^{(2)}_{\chpt}
=\frac{F_0^2}{4}{\tr}\left[D_\mu U (D^\mu U)^\dagger \right]
+ \frac{F_0^2}{4}{\tr} \left[\chi U^\dagger +U\chi^\dagger \right],
\end{align}
where $F_0$ is the pion decay constant in the chiral limit and $U$ is related to the pNGBs by
\begin{align}
U=\text{exp}\left[i{\sqrt{2} \Phi \over F_0}\right], \quad
\Phi & = 
\begin{pmatrix}
\frac{\pi^0}{\sqrt{2}}+\frac{\eta}{\sqrt{6}} & \pi^+ & K^+
\\
\pi^- & -\frac{\pi^0}{\sqrt{2}}+\frac{\eta}{\sqrt{6}} & K^0
\\
K^- & \bar{K}^0 & -\sqrt{\frac{2}{3}}\eta
\end{pmatrix},
\end{align}
and the covariant derivative of $U$ and the combined scalar source $\chi$ are 
\begin{align}
D_\mu U = \partial_\mu U-i l_\mu U+i U r_\mu,
\quad 
\chi = 2 B(s-ip).
\end{align}
The tensor source first appears at $\calO(p^4)$ \cite{Cata:2007ns}, 
which will yield the DM EM dipole moments. The relevant term is
\begin{align}
\label{eq:chilag:p4}
\calL_{\chi {\rm PT}}^{(4)} 
\supset \Lambda_1 {\tr} \left[ t_+^{\mu\nu}  f_{+\mu\nu}\right],
\end{align}
where $\Lambda_1$ is a low energy constant (LEC) which is usually parameterized
in terms of the chiral symmetry breaking scale $\Lambda_\chi$ in the form, 
$\Lambda_1 =c_T \Lambda_\chi/(16\pi^2)$, with $c_T$ an unknown dimensionless constant.
Here the tensor field matrices in flavor space are given by
\begin{align}
t^{\mu\nu}_+  = u^\dagger t^{\mu\nu} u^\dagger +u t^{\mu\nu\dagger}u,
\quad
f^{\mu\nu}_+  =  u F^{\mu\nu}_L u^\dagger + u^\dagger F^{\mu\nu}_R u,
\end{align}
with $u^2=U$. 
The chiral field strength tensors read
\begin{align}
F^{\mu\nu}_{L} = \partial^\mu l^\nu-\partial^\nu l^\mu-i[l^\mu,l^\nu],
\quad
F^{\mu\nu}_{R} = \partial^\mu r^\nu-\partial^\nu r^\mu-i[r^\mu,r^\nu].
\end{align}
For our purpose, the vector external sources are recognized as,
\begin{align}
l_\mu=r_\mu= -e A_\mu \,{\rm diag}(Q_u, Q_d, Q_s),
\end{align}
where $A_\mu$ denotes the photon field and $Q_q$ the electric charge of quark $q$ in units of $e>0$.

Expanding \cref{eq:chilag:p4} to the lowest order in pNGB fields, 
the following DM edm and mdm interactions arise,
\begin{align}
\calL_{\chi {\rm PT}}^{(4)} \supset 
{\mu_\chi \over 2} (\bar \chi \sigma^{\mu\nu} \chi) F_{\mu\nu}
+ {d_\chi \over 2 } (\bar \chi i \sigma^{\mu\nu}\gamma_5 \chi) F_{\mu\nu},
\end{align}
with the DM mdm and edm being
\begin{subequations}
\label{eq:di-mom-coe}
\begin{align}
\mu_\chi & = 
- {e c_T \Lambda_\chi  \over 12 \pi^2} \left( \sum_q 3 Q_q C_{\chi q}^{\tt T1} m_q \right)
= {e c_T \Lambda_\chi  \over 12 \pi^2} ( C_{\chi d}^{\tt T1} m_d - 2 C_{\chi u}^{\tt T1} m_u
+ C_{\chi s}^{\tt T1} m_s ),  
\\
d_\chi & =  - {e c_T \Lambda_\chi  \over 12 \pi^2} \left( \sum_q 3 Q_q  C_{\chi q}^{\tt T2} m_q \right)
= {e c_T \Lambda_\chi  \over 12 \pi^2} ( C_{\chi d}^{\tt T2} m_d - 2 C_{\chi u}^{\tt T2} m_u
+  C_{\chi s}^{\tt T2} m_s ). 
\end{align}
\end{subequations}
Assuming the DM EM dipole moments are dominated by these nonperturbative contributions,
we can establish the relation between the scale 
$\Lambda = |C_{\chi q}^{\tt T1,2}|^{-1/3}$
for the DM-quark tensor operators and the dipole moments via
\begin{subequations}
\label{eq:DMEM}
\label{eq:di-mom-Lambda}
\begin{align}
\Lambda & =  \left| {e c_T \Lambda_\chi  (3 Q_q m_q) \over 12 \pi^2}{1 \over \mu_\chi} \right|^{1/3}
\approx 4\,{\rm GeV} \left| {3 Q_q m_q \over 2\,{\rm MeV}} { 10^{-9} \mu_B \over \mu_\chi } \right|^{1/3}, 
\label{eq:MDM}
\\
\Lambda & =  \left| {e c_T \Lambda_\chi  (3 Q_q m_q) \over 12 \pi^2}{1 \over d_\chi} \right|^{1/3}
\approx 50\,{\rm GeV} \left| {3 Q_q m_q \over 2\,{\rm MeV}} {10^{-23} {\rm e\, cm} \over d_\chi } \right|^{1/3},
\label{eq:EDM}
\end{align}
\end{subequations}
where only one flavor quark contribution is assumed. 
In the above numerical illustration, we have used the model estimation of the LEC constant $c_T = -3.2$ in \cite{Mateu:2007tr}; 
there are also other studies that give a smaller magnitude, $c_T\approx -1.0(2)$ \cite{Baum:2011rm,Ecker:1988te},
which will reduce $\Lambda$ in \cref{eq:DMEM} by a factor of 0.7.
For the flavor symmetric case with 
$C_{\chi u}^{\tt T1,2}=C_{\chi d}^{\tt T1,2}=C_{\chi s}^{\tt T1,2}$, the $\mu_\chi$ and $d_\chi$ are 
totally dominated by the strange quark while the contributions from up and down quarks almost cancel out due to the approximate mass relation, $Q_um_u+Q_dm_d\approx 0$. 

Now we turn to the matching onto DM-nucleon interactions. 
First of all, only the single-nucleon currents are present at LO in chiral power counting. 
Thus, we neglect the subleading contribution from higher chiral power terms and two-nucleon 
current \cite{Hoferichter:2018acd}. 
The nucleon matrix element of the DM-quark operators can be parameterized in terms of form factors which are restricted by 
Lorentz covariance, discrete symmetries, and algebraic identities for Dirac matrices and spinors. 
For the tensor operator, there are three form factors \cite{Adler:1975he}, 
{\small
\begin{align}
\label{eq:ts:ff}
\langle N(k_2)|\bar q \sigma^{\mu\nu} q|N(k_1)\rangle = 
\bar{u}_{k_2}\left[F_{T,0}^{q/N}(q^2)\sigma^{\mu\nu}
+F_{T,1}^{q/N}(q^2) \frac{i \gamma^{[\mu}q^{\nu]} }{m_N} 
+F_{T,2}^{q/N}(q^2) \frac{i k_{12}^{[\mu} q^{\nu]} }{m_N^2}  \right] u_{k_1},
\end{align} 
where $q^\mu=k_2^\mu-k_1^\mu$, $k_{12}^\mu=k_1^\mu+k_2^\mu$, and $m_N$ is the nucleon
mass.\footnote{The above parameterization is slightly different from the one used in the package {\tt DirectDM} 
\cite{Bishara:2017pfq,Bishara:2017nnn}, 
\begin{align*}
\langle N(k_2)|m_q \bar q \sigma^{\mu\nu} q|N(k_1)\rangle%_{\tt DirectDM}
= 
\bar{u}_{k_2}\left[F_{T,0}^{q/N}(q^2)\sigma^{\mu\nu}
+F_{T,1}^{q/N}(q^2) \frac{i \gamma^{[\mu}q^{\nu]} }{2 m_N} 
+F_{T,2}^{q/N}(q^2) \frac{i q^{[\mu} k^{\nu]}_{12} }{m_N^2}  \right] u_{k_1}.
\end{align*}
}
In the B$\chi$PT framework, the form factors $F_{T,i}^{q/N}(q^2)$ ($i=0,1,2$) are calculated order by order in the chiral expansion. 
Owing to the absence of light pseudoscalar poles as well as 
the small momentum squared of our interest ($|q^2| \sim \calO(1\,\rm MeV^2)$), 
the form factors can be Taylor-expanded around $q^2=0$, 
with the largest contributions coming from their values evaluated at $q^2=0$.
For the tensor charges, $F_{T,0}^{q/N}(0)=g_T^{q/N}$, we will use the lattice QCD result in \cite{Gupta:2018lvp},\footnote{
\label{foot:gTs}
We find different values of $g_T^{s/N}$ are quoted in the literature.
There are two independent lattice calculations yielding consistent results, 
with $g_T^{s/N} = -0.0027(16)$
in \cite{Gupta:2018lvp}
and $g_T^{s/N} = -3.19 \times 10^{-3}(69)(2)(22)$ in
the erratum to \cite{Alexandrou:2017qyt} 
(correcting its first-version value $g_T^{s/N} = -3.2\times 10^{-4}(24)(0)$).
Subsequent quotations made typos;
e.g., Refs.\,\cite{flavourLatticeAveragingGroup:2019iem,DelNobile:2021wmp} quoted $g_T^{s/N} = -0.027(16)$ (see also footnote 5) while Ref.\,\cite{Bishara:2017pfq} and the package \href{https://directdm.github.io}{\tt DirectDM} quoted $g_T^{s/N}= (3.2 \pm 8.6)\times 10^{-4}$. 
The tensor charge $g_T^{s/N}$ is also denoted by $\delta_s^{N}$ in the literature.  
}
\begin{align}
F_{T,0}^{u/p}(0) 
&= 0.784\,(28)(10), &
F_{T,0}^{d/p}(0) 
&= -0.204\,(11)(10), &
F_{T,0}^{s/p}  
& = -0.0027\,(16),
\end{align}
while for the other two we will adopt the recent result obtained in \cite{Hoferichter:2018zwu}, 
\begin{subequations}
\begin{align}
F_{T,1}^{u/p}(0)
&=  -1.5\,(1.0), &
F_{T,1}^{d/p}(0)  
& = 0.5\,(3) , &
F_{T,1}^{s/p}(0) 
& = 0.009\,(5),
\\
F_{T,2}^{u/p}(0) 
&=  0.1\,(2), &
F_{T,2}^{d/p}(0)  
&= -0.6\,(3), &
F_{T,2}^{s/p}(0) 
&= -0.004\,(3),
\end{align}
\end{subequations}
where isospin symmetry is implied, i.e., $F_{T,i}^{u/p}=F_{T,i}^{d/n}$, $F_{T,i}^{d/p}=F_{T,i}^{u/n}$, and 
$F_{T,i}^{s/p}=F_{T,i}^{s/n}$. 
An alternative estimation based on hadronic models is given in \cite{Pasquini:2005dk} for the form factors $F_{T,1}^{q/N}$, whose values (including vanishing central value for $q=s$) are usually employed in the DMDD community \cite{Bishara:2017pfq,Bishara:2017nnn}.

To calculate the DM-nucleus matrix element, 
nonrelativistic (NR)\footnote{Note the slight font difference for this abbreviation ``NR'' and the ``$\NR$'' used for ``nuclear recoil''.}
reduction of the nucleon-level amplitude should be performed to make connections with the treatment in nuclear many-body methods.
This is done by taking both the DM and nucleon spinors to the NR limit 
and expressing the amplitude as a combination of various NR quantities.
The NR amplitude can be equivalently obtained using NR operators.
In this operator language,
according to \cref{eq:ts:ff},
the chiral LO NR expansions of the contact tensor operators are given by,
\begin{subequations}
\label{eq:tensorNR}
\begin{align}
C_{\chi q}^{\tt T1} \OTa  
\overset{\rm NR}{\rightarrow}
&\,\,
32 C_{\chi q}^{\tt T1} m_q  F_{T,0}^{q/N}  m_\chi m_N \calO_4^N,
\\
C_{\chi q}^{\tt T2} \OTb 
\overset{\rm NR}{\rightarrow} 
&\,\,
8 C_{\chi q}^{\tt T2} m_q\left[ m_N F_{T,0}^{q/N} \calO_{10}^N 
- m_\chi \left(F_{T,0}^{q/N}-2 F_{T,1}^{q/N}- 4 F_{T,2}^{q/N} \right)\calO_{11}^N
\right.
\nonumber
\\
&\left.
- 4 m_\chi m_N F_{T,0}^{q/N} \calO_{12}^N \right],
\end{align}
\end{subequations}
In the NR reduction of $\OTb$, we have included contributions from the $F_{T,2}^{q/N}$ term, which is spin-independent and potentially comparable to other terms. Including this term is essential for consistency in our analysis, and to the best of our knowledge, it has been ignored in previous calculations \cite{Bishara:2017nnn}. 
For the mdm and edm interactions, the NR expansions read 
{\small
\begin{subequations}
\label{eq:dipoleNR}
\begin{align}
\frac{\mu_\chi}{2}(\bar\chi \sigma^{\mu\nu}\chi) F_{\mu\nu}  
&\overset{\rm NR}{\rightarrow}
-2 e\mu_\chi\left[ m_N Q_N \calO_1^N 
+ 4\frac{m_\chi m_N}{\pmb{q}^2}Q_N \calO_5^N
+ 2 m_\chi g_N \left(\calO_4^N
-\frac{\calO_6^N}{\pmb{q}^2}\right)\right],
\\
\frac{d_\chi}{2}(\bar\chi i\sigma^{\mu\nu}\gamma_5\chi) F_{\mu\nu}  
&\overset{\rm NR}{\rightarrow}
-8 \frac{m_\chi m_N}{\pmb{q}^2} e d_\chi Q_N \calO_{11}^N,
\end{align}
\end{subequations} 
}\noindent
where $Q_N$ represents the nucleon electric charge in units of $e$, and $g_N$ is
the nucleon Land\'{e} $g$-factor, 
with $g_p=5.59$ and $g_n=-3.83$ for the proton and neutron respectively.
The involved NR operators in \cref{eq:tensorNR} and \cref{eq:dipoleNR}
are \cite{DelNobile:2018dfg,Fitzpatrick:2012ix,Anand:2013yka}
\begin{align}
\nonumber
\calO_1^N &\equiv \mathbb{1}_\chi \mathbb{1}_N, &
\calO_4^N &\equiv \pmb{S}_\chi \cdot \pmb{S}_N, &
\calO_5^N &\equiv 
i \pmb{S}_\chi \cdot (\pmb{q}\times\pmb{v}_N^\perp), &
\calO_6^N & \equiv 
(\pmb{S}_\chi\cdot\pmb{q}) (\pmb{S}_N\cdot\pmb{q}), & 
\\
\calO_{10}^N &\equiv i\pmb{S}_N\cdot\pmb{q}, &
\calO_{11}^N & \equiv i\pmb{S}_\chi\cdot\pmb{q}, & 
\calO_{12}^N & \equiv 
\pmb{v}_N^\perp\cdot(\pmb{S}_\chi\times \pmb{S}_N), &
\label{eq:NRope}
\end{align}
where $\pmb{q}$ is the three-momentum transfer, 
and $\pmb{S}_\chi$ and $\pmb{S}_N$ are respectively the DM and nucleon spin operator. 
Here the ``elastic'' transverse velocity is defined by
\begin{align}
\pmb{v}^\perp_N \equiv
\pmb{v}_{\chi}-\frac{\pmb{q}}{2\mu_{N\chi}},
\end{align}
with $\pmb{v}_{\chi}$ being the incoming DM-nucleon relative velocity
and $\mu_{N\chi}$ the reduced mass for the DM-nucleon system.

%%%%%%%%%%%%%%%%%%%%%%%%%%%%%%
\section{XENON1T constraints}
\label{sec:xenon1}
%%%%%%%%%%%%%%%%%%%%%%%%%%%%%%

As mentioned in \cref{sec:match}, $\OTa$ and $\OTb$
can induce not only 4-fermion DM-nucleon interactions
but also the EM dipole moments of the DM.
Hereafter, we denote their contribution in DMDD experiments as the short-distance (SD) and long-distance (LD) one respectively.
In previous calculations of DMDD constraints on $\OTa$ and $\OTb$ \cite{Kang:2018odb,Kang:2018rad,Tomar:2022ofh},
only the DM-nucleus scattering induced by the SD contribution is considered.
A consistent calculation needs to take into account both the SD and LD contributions at the amplitude level,
where interference effect between the two is generally expected.
In addition to DM-nucleus scattering, the LD dipole operators can also induce DM-electron scattering,
as shown in \cref{fig:Feyn-dia1}. Due to the excellent potential of DM-electron scattering
to probe low-mass DM \cite{Essig:2011nj,Essig:2012yx}, significant improvements
for the constraints in the low-mass region are possible
if we take into account the DM-electron scattering induced by the LD dipole contribution.
For the operators with vector or axial-vector currents, only the SD contributions to the DMDD are induced and the probed DM mass region is above a few GeV from the NR signals
and only extends to 0.1\,GeV or so from the Migdal effect 
 \cite{Tomar:2022ofh}.
However, we will show that the LD contribution from the tensor operators is capable of probing DM as low as 5 MeV.

%%%%
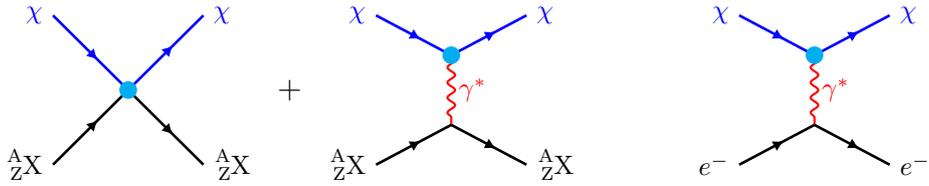
\begin{figure}[t]
\centering
\begin{tikzpicture}[mystyle,scale=1.1]
\begin{scope}[shift={(1,1)}]3
\draw[f,blue] (-1.5,1.5) node[left]{$\chi$}  -- (0,0);
\draw[f, blue] (0, 0) -- (1.5,1.5) node[right]{$\chi$};
\draw[f] (-1.5,-1.5) node[left]{$\ce{ ^A_ZX}$}-- (0,0);
\draw[f] (0, 0) -- (1.5,-1.5) node[right]{$\ce{ ^A_ZX}$};
\draw[draw=cyan,fill=cyan] (0,0) circle (0.15cm);
\end{scope}
\end{tikzpicture}
\begin{tikzpicture}[mystyle,scale=1.1]
\draw[] (0,1) node[]{\large +};
\end{tikzpicture}
\begin{tikzpicture}[mystyle,scale=1.1]
\begin{scope}[shift={(1,1)}]3
\draw[f,blue] (-1.5,1.5) node[left]{$\chi$}  -- (0,0.7);
\draw[f, blue] (0, 0.7) -- (1.5,1.5) node[right]{$\chi$};
\draw[v, red, thick] (0,0.7) -- (0,-0.7) node[midway,xshift = 8 pt]{$\gamma^*$};
\draw[f] (-1.5,-1.5) node[left]{$\ce{ ^A_ZX}$}-- (0,-0.7);
\draw[f] (0, -0.7) -- (1.5,-1.5) node[right]{$\ce{ ^A_ZX}$};
\draw[draw=cyan,fill=cyan] (0,0.7) circle (0.15cm);
\end{scope}
\end{tikzpicture}\quad \quad\quad \quad
\begin{tikzpicture}[mystyle,scale=1.1]
\begin{scope}[shift={(1,1)}]3
\draw[f,blue] (-1.5,1.5) node[left]{$\chi$}  -- (0,0.7);
\draw[f, blue] (0, 0.7) -- (1.5,1.5) node[right]{$\chi$};
\draw[v, red, thick] (0,0.7) -- (0,-0.7) node[midway,xshift = 8 pt]{$\gamma^*$};
\draw[f] (-1.5,-1.5) node[left]{$e^-$}-- (0,-0.7);
\draw[f] (0, -0.7) -- (1.5,-1.5) node[right]{$e^-$};
\draw[draw=cyan,fill=cyan] (0,0.7) circle (0.15cm);
\end{scope}
\end{tikzpicture}
\caption{Feynman diagrams for DM-nucleus scattering (the left two diagrams)
and DM-electron scattering (the rightmost diagram) induced by $\OTa$ and $\OTb$.}
\label{fig:Feyn-dia1}
\end{figure}
%%%%

The XENON1T experiment is a DMDD experiment with a dual-phase time projection chamber, 
in which both DM-electron and DM-nucleus scattering can induce prompt scintillation photons (S1 signal)
and drift electrons (S2 signal) \cite{XENON:2018voc}. 
According to the strength ratio between the S1 and S2 signals,
 the $\NR$ and  $\ER$  signals can be well distinguished \cite{XENON:2018voc}.
Comprehensive searches for DM particles have been carried out 
by the XENON1T collaboration, including $\NR$ signals for the DM-nucleus scattering \cite{XENON:2018voc}, 
$\ER$ signals for the DM-electron scattering \cite{XENON:2021qze}, 
and S2-only signals for the Migdal effect \cite{XENON:2019zpr}.
Hereafter, we denote the corresponding constraints from the above three types of signals
as the $\NR$, $\ER$, and Migdal constraints respectively. In this section, we utilize the XENON1T experiment as a benchmark to recalculate the constraints on $\OTa$ and $\OTb$
with the data given in \cite{XENON:2018voc,XENON:2019zpr,XENON:2021qze},
in which we consistently take into account both the SD and LD contributions.

Our analysis will cover the cases in which the flavor SU(3) symmetry is imposed or not.
For the flavor conserving case, a universal Wilson coefficient
is assumed for the operators $\OTa$ and $\OTb$ with all of $u,d,s$ quarks. This is the 
case usually adopted in the literature. We will also study the separate contributions from individual quarks without assuming the flavor symmetry. We will denote the 90\,\% confidence level (C.L.) constraint on $\Lambda$ as $\Lambda_q$ in the flavor conserving case and $\Lambda_{u,d,s}$ when the three light quarks are treated separately. The effective tensor interactions may also appear without being accompanied by a quark mass; an example of how this occurs in a UV model will be shown in Section \ref{sec:nonDD}. In that case the Wilson coefficients of the dimension-six tensor operators are parameterized by an effective scale $\tilde\Lambda$, and the upper bound on $\Lambda$ will be translated to that on $\tilde\Lambda=\Lambda\sqrt{\Lambda/m_q}$. In \cref{fig:Xenon1T-T1,fig:Xenon1T-T2}, we will show the constraints on both $\Lambda$ and $\tilde{\Lambda}$ for individual quark contributions.

%%%%%%%%%%%%%%%%%%%%%%%%%%%%%%
\subsection{DM-nucleus scattering cross section}
%%%%%%%%%%%%%%%%%%%%%%%%%%%%%%

In this subsection,
we give the differential cross section for the DM-nucleus scattering 
from the two tensor interactions with both the SD and LD 
contributions included. 
The distribution with respect to  
the nuclear recoil energy $E_R$ ($E_R=\pmb{q}^2/2m_A$)
in the NR limit is given by \cite{DelNobile:2021wmp}
\begin{align}
\frac{d\sigma_T}{d E_R}{=} 
\frac{1}{32 \pi}\frac{1}{m_\chi^2 m_A}\frac{1}{v^2} \overline{|\mathcal{M}|^2},
\end{align}
where $m_A$ is the mass of the target nucleus, 
$v$ is the speed of the incoming DM particle in the lab frame,
and $\overline{|\mathcal{M}|^2}$
is the amplitude squared that has been averaged/summed over the initial/final spin states. 
The DM-nucleus amplitude $\mathcal{M}$ is given by the sum over all protons and neutrons
in the nucleus of the single-nucleon amplitude derived in \cref{sec:match}. 
Furthermore, the corresponding NR operators are decomposed into spherical components 
with a definite angular momentum, which is suitable to computations 
for a nucleus in an eigenstate of the total angular momentum. 
By performing a multipole expansion, 
the unpolarized amplitude squared can be represented in a compact form \cite{DelNobile:2021wmp},
\begin{align}
\label{eq:samp}
\overline{|\mathcal{M}|^2} =
\frac{m_A^2}{m_N^2} \sum_{i,j} \sum_{N,N^\prime=p,n} f_i^N (\pmb{q}^2) f_j^{N^\prime} (\pmb{q}^2) 
F_{i,j} ^{(N,N^\prime)}(\pmb{q}^2,\pmb{v}^{\perp 2}_{T}),
\end{align}
where $i$ and $j$ span the NR operator basis. The squared form factors
$F_{i,j} ^{(N,N^\prime)}(\pmb{q}^2,\pmb{v}^{\perp 2}_{T})$ depend on the nuclear responses
as well as $\pmb{q}^2$ and $\pmb{v}^{\perp 2}_{T}\equiv v^2-v_{\rm min}^2$,
and the relevant ones in our consideration are collected in \cref{app:sform}.
Here $v_{\rm min} = \sqrt{E_R m_A}/({\sqrt{2}\mu_{A\chi}})$
is the minimum velocity for DM to
induce a nuclear recoil energy $E_R$ with $\mu_{A\chi}$ 
being the reduced mass for the DM-nucleus system.
Detailed formulations of these squared form factors for various nuclei 
have been provided in \cite{Fitzpatrick:2012ix}. 

The functions $f_i^N(\pmb{q}^2)$ are determined by particle physics from the ultraviolet down to the chiral scale. 
In our case, they are contributed by both DM-nucleon and DM dipole interactions induced
from the tensor DM-quark operators. Concretely, for the operator $\OTa$, the functions that do not vanish are as follows:
\begin{subequations}
\label{eq:matOta}
\begin{align}
f_1^N &= -2 e\mu_\chi m_N Q_N,
\\
f_4^N &=-4e\mu_\chi m_\chi g_N +\sum_{q=u,d,s} 32 C_{\chi q}^{\tt T1} m_q F_{T,0}^{q/N} m_\chi m_N,
\\
f_5^N &= -8 e\mu_\chi\frac{m_\chi m_N}{\pmb{q}^2} Q_N,
\\
f_6^N &= 4 e\mu_\chi\frac{m_\chi}{\pmb{q}^2} g_N.
\end{align} 
\end{subequations}
The inclusion of the DM mdm ($\mu_\chi$) has resulted in a new term in $f_4^N$ and nonvanishing $f_{1,5,6}^N$. 
The mdm term in $f_{4,5,6}^N$ will interfere with the usual SD term in $f_4^N$ in the amplitude squared,
while $f_1^N$ being DM spin independent will not interfere with $f_{4,5,6}^N$.
Similarly, for the operator $\OTb$, the nonvanishing functions are
\begin{subequations}
\label{eq:matOtb}
\begin{align}
f_{10}^N &= \sum_{q=u,d,s} 8 C_{\chi q}^{\tt T2} m_q F_{T,0}^{q/N} m_N,
\\
f_{11}^N &= -8\frac{m_\chi m_N}{\pmb{q}^2}e d_\chi Q_N 
-\sum_{q=u,d,s} 8 C_{\chi q}^{\tt T2} m_q(F_{T,0}^{q/N}-2 F_{T,1}^{q/N}-4 F_{T,2}^{q/N}) m_\chi,
\\
f_{12}^N &= -\sum_{q=u,d,s} 32 C_{\chi q}^{\tt T2} m_q F_{T,0}^{q/N} m_\chi m_N.
\end{align}
\end{subequations}
The DM edm leads to an additional term in $f_{11}^N$,
which will interfere with the usual SD term in
$f_{11}^N$ and $f_{12}^N$ but not with $f_{10}^N$ as the latter is independent of the DM spin. 

%%%%%%%%%%%%%%%%%%%%%%%%%%%%%%
\subsection{Constraint from $\NR$ signals}
%%%%%%%%%%%%%%%%%%%%%%%%%%%%%%
The differential event rate for   $\NR$   signals is given by,
\begin{align}
\frac{dR_{\tt NR}}{ dE_R} = 
\frac{\rho_\chi}{m_\chi}{1\over m_A} \int_{v_{\rm min}(E_R)}^{v_{\rm max}} dv F(v) v
{d\sigma_T\over dE_R} (v,E_R),
\label{eq:NR}
\end{align}
where $\rho_\chi = 0.3\,\si{GeV /cm^3}$ is the local DM energy density near the Earth
and $F(v)$ is the DM velocity distribution in the lab frame. In the actual calculation,
the total rate is a sum of contributions from each isotope weighted by its mass fraction in the nuclear target.
Note that the angular distribution of the DM velocity has been integrated out in $F(v)$,
since the target nuclei are considered to be at rest and unpolarized in the lab frame.

In the galaxy rest frame, the DM velocity obeys a normal Maxwell-Boltzmann distribution
with the circular velocity $v_0= 220\,\rm km/s$ \cite{Smith:2006ym}, 
which leads to \cite{Lewin:1995rx,DelNobile:2021wmp}
\begin{align}
F(v)= 
\frac{v}{\sqrt{\pi} v_0 v_E}
\begin{cases}
 e^{-(v-v_E)^2 / v_0^2}-e^{-(v+v_E)^2 / v_0^2}, 
& \text{for}~ 0 \leq v \leq v_{\mathrm{esc}}-v_E 
\vspace{0.1cm}
\\ 
 e^{-(v-v_E)^2 / v_0^2}-e^{-v_{\mathrm{esc}}^2 / v_0^2}, 
& \text{for~}v_{\mathrm{esc}}-v_E<v \leq v_{\mathrm{esc}}+v_E
\end{cases}.
\end{align}
Here we adopt the averaged Earth relative velocity $v_E= 232\,\rm km/s$ \cite{Wang:2021oha}
and the escape velocity $v_{\rm esc} = 544\,\rm km/s$ \cite{Smith:2006ym},
which leads to the maximal DM velocity in the lab frame being
$v_{\rm max}=v_{\rm esc}+v_E=776\,\rm km/s$.

We calculate the constraint based on the $\NR$  events given in \cite{XENON:2018voc},
for an exposure of $w=$ 1.0 ton-yr.
Taking into account the SM backgrounds,
the 90\% C.L. constraint is obtained by the criterion of $N_{\tt NR}^s < 7$ \cite{Kang:2018rad}, 
where the number of $\NR$ events induced by DM-nucleus scattering is calculated by 
\begin{align}
N_{\tt NR}^s = 
 w \int_0^{70\,\rm keV} \epsilon_{\tt NR} (E_R) \frac{dR_{\tt NR}}{d E_R} dE_R.
\end{align}
Here we adopt the $\NR$ signal efficiency, $\epsilon_{\tt NR}(E_R)$, 
given in Fig.\,(1) of \cite{XENON:2018voc}.

The XENON1T constraints on $\OTa$ 
and $\OTb$ from $\NR$ are shown as red curves in \cref{fig:Xenon1T-T1} and \cref{fig:Xenon1T-T2}.
The SD contribution (red dashed curves) dominates over the LD contribution
in the constraints on $\OTa$ for the valence $u$ and $d$ quarks.
For the sea $s$ quark, the SD contribution is relatively
less important than the LD one (red dotted curves) for small $m_\chi$
until $m_\chi \gtrsim 50\,\rm GeV$ when their constructive
interference starts to become significant.
Regarding the constraints on $\OTb$, 
the LD contribution dominates overwhelmingly for the $s$ quark, 
while the $u$ and $d$ quarks exhibit comparable but varying
contributions from the SD and LD mechanisms. Note that the interference effect is 
constructive (destructive) in the $u$ ($d$) quark scenario.
This distinct behavior is due to the charge sign difference between the $u$ and $d$ quarks.
For both operators $\OTa$ and $\OTb$ in the flavor conserving case,  
the constraint from SD is dominated by the $d$ quark contribution 
while the LD constraint is dominated by the $s$ quark contribution. 
Especially for $\OTb$, the LD contribution always dominates, 
but becomes comparable with the SD one for a large $m_\chi$
where a significant destructive interference pattern is evident in the full constraint (red solid curve).

\begin{figure}%[htbp]
\includegraphics[width=0.5 \textwidth]{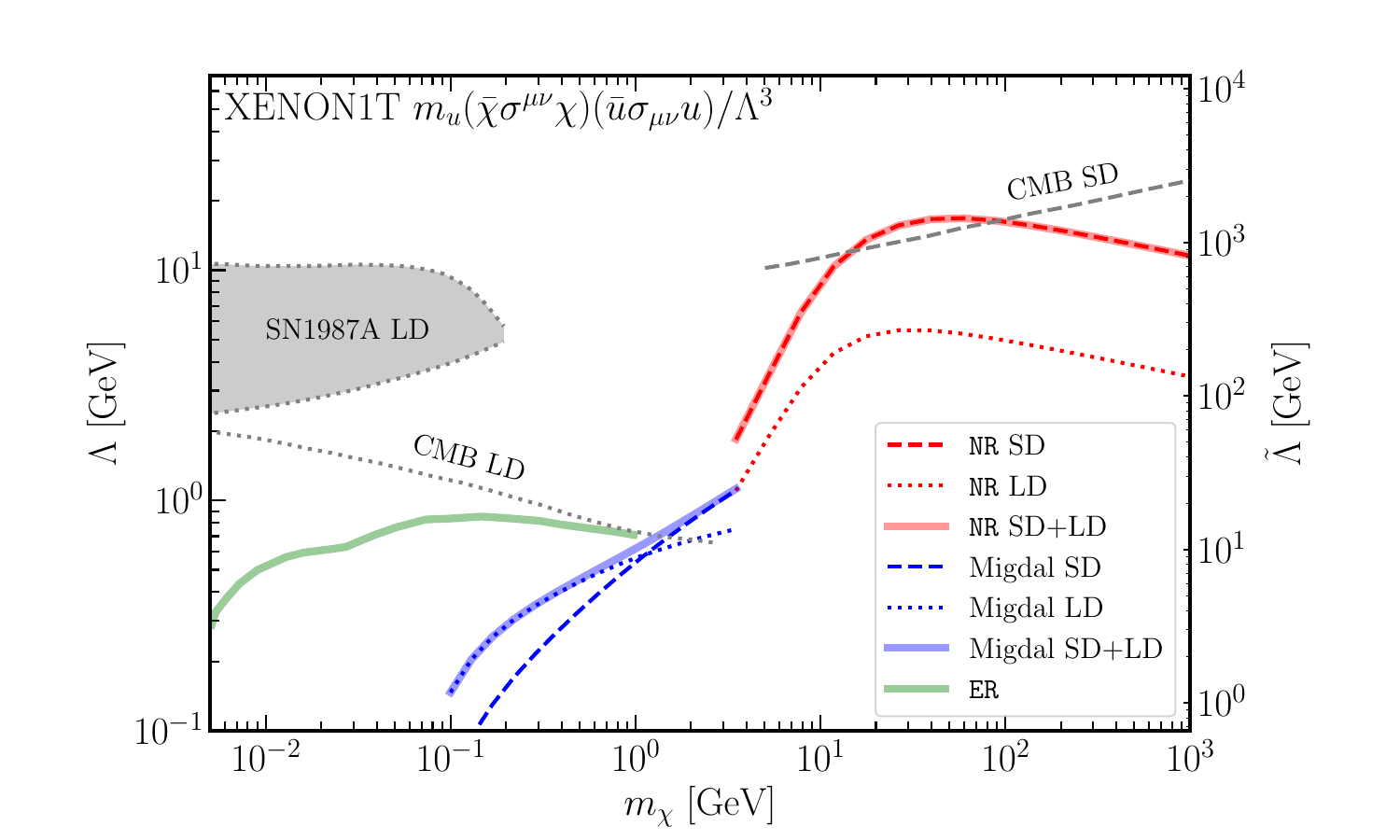}
\includegraphics[width=0.5 \textwidth]{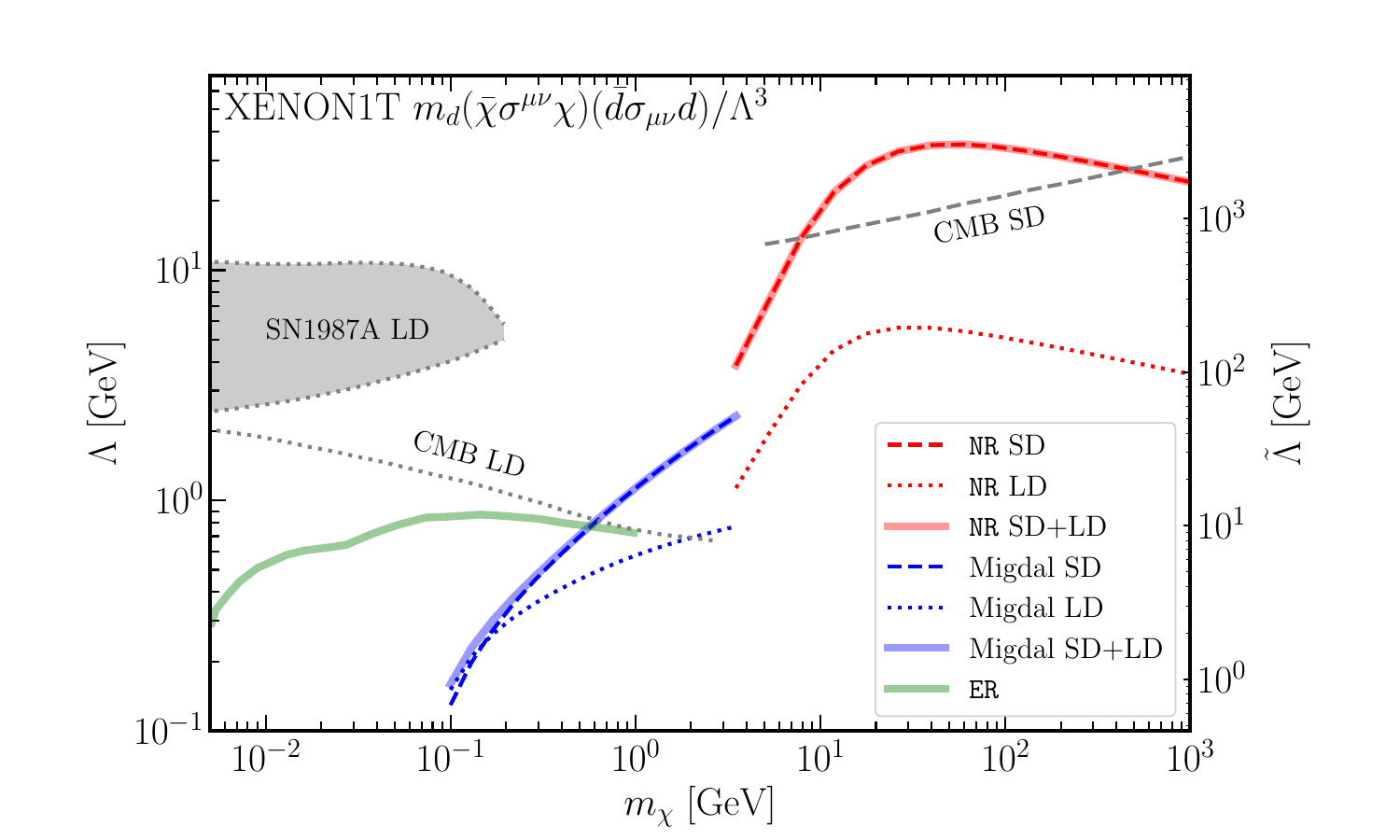}
\\
\includegraphics[width=0.5\textwidth]{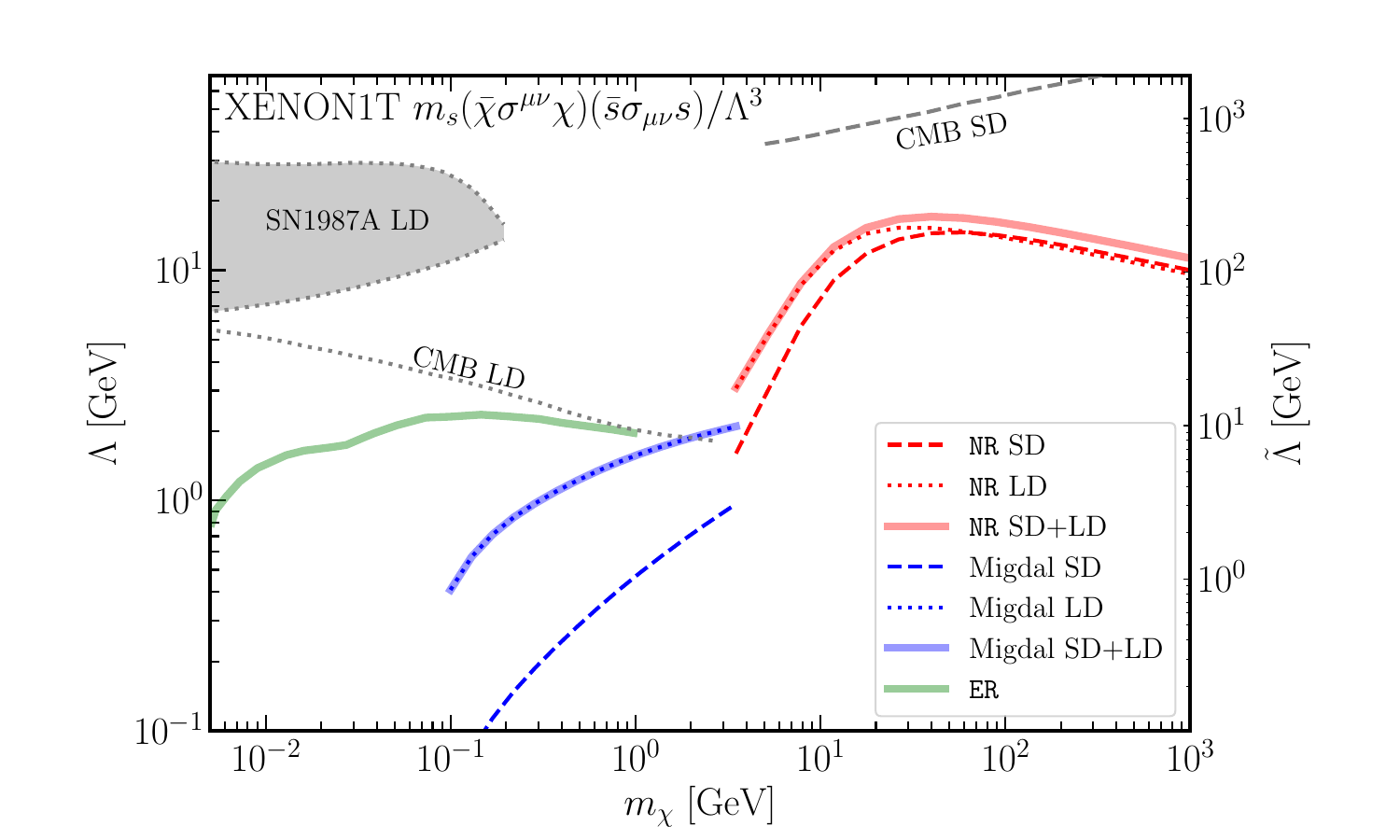}
\includegraphics[width=0.5 \textwidth]{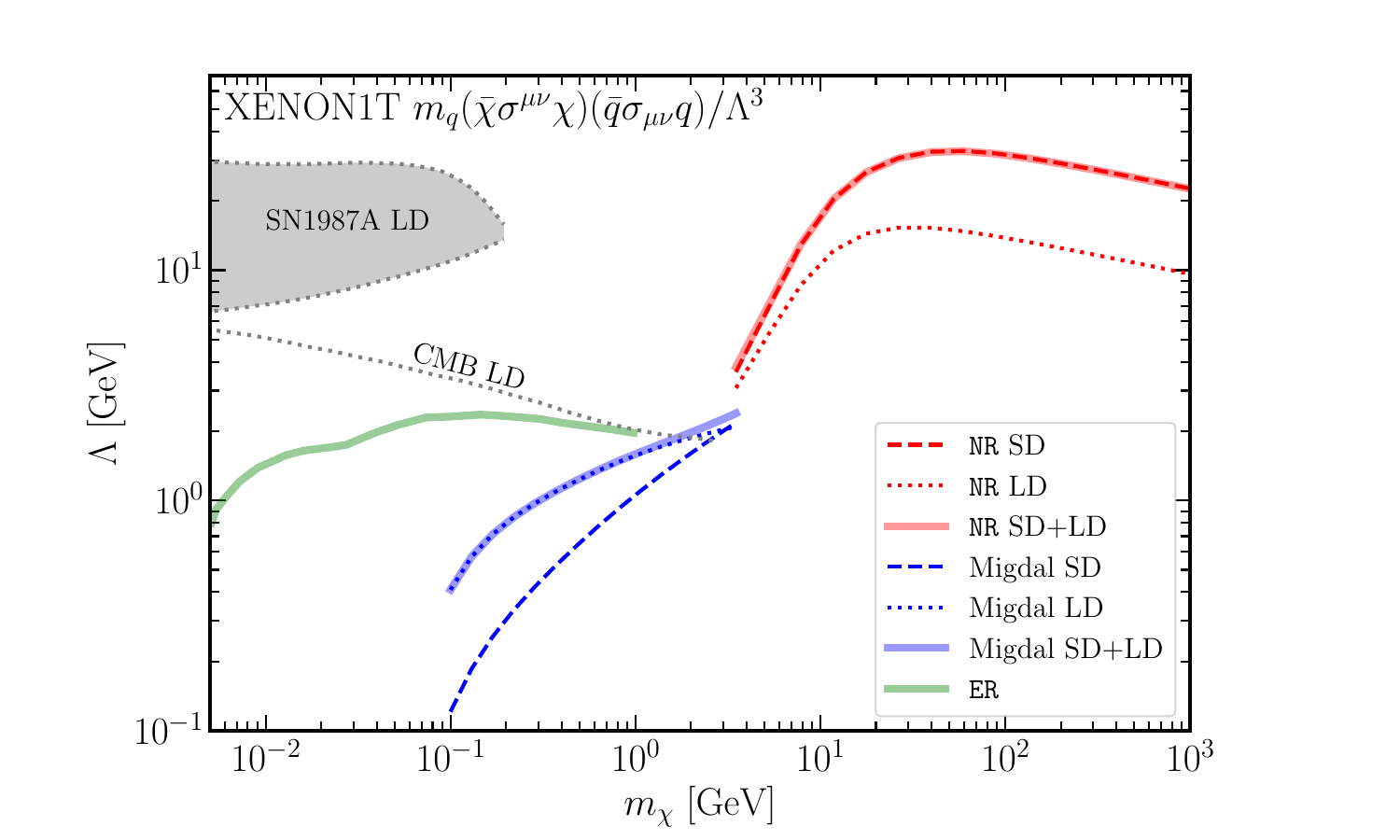}
\vspace{-5mm}
\caption{
XENON1T constraints on $\OTa$ from $\NR$ signals (red), the Migdal effect (blue), and $\ER$ signals (green). Here $\Lambda$ denotes the effective scale associated with the dimension-seven $\OTa$ and $\tilde\Lambda$ for the corresponding dimension-six operator with the quark mass $m_q$ removed. 
For the $\NR$ and Migdal effect cases, we consider the constraints with SD-only (dashed), LD-only (dotted), and full (solid) contributions respectively. For the $\ER$ case, only the LD contribution is involved. In all panels excluding the bottom-right one, the individual contribution from the $u$, $d$, and $s$ quark is taken into account separately. In the bottom-right panel, a flavor universal coupling is assumed for all of the $u$, $d$, and $s$ quarks. 
The constraints from SN1987A and CMB discussed in \ref{sec:nonDD} are also shown here. 
}
\label{fig:Xenon1T-T1}
\end{figure}

\begin{figure}%[htbp]
\includegraphics[width=0.5 \textwidth]{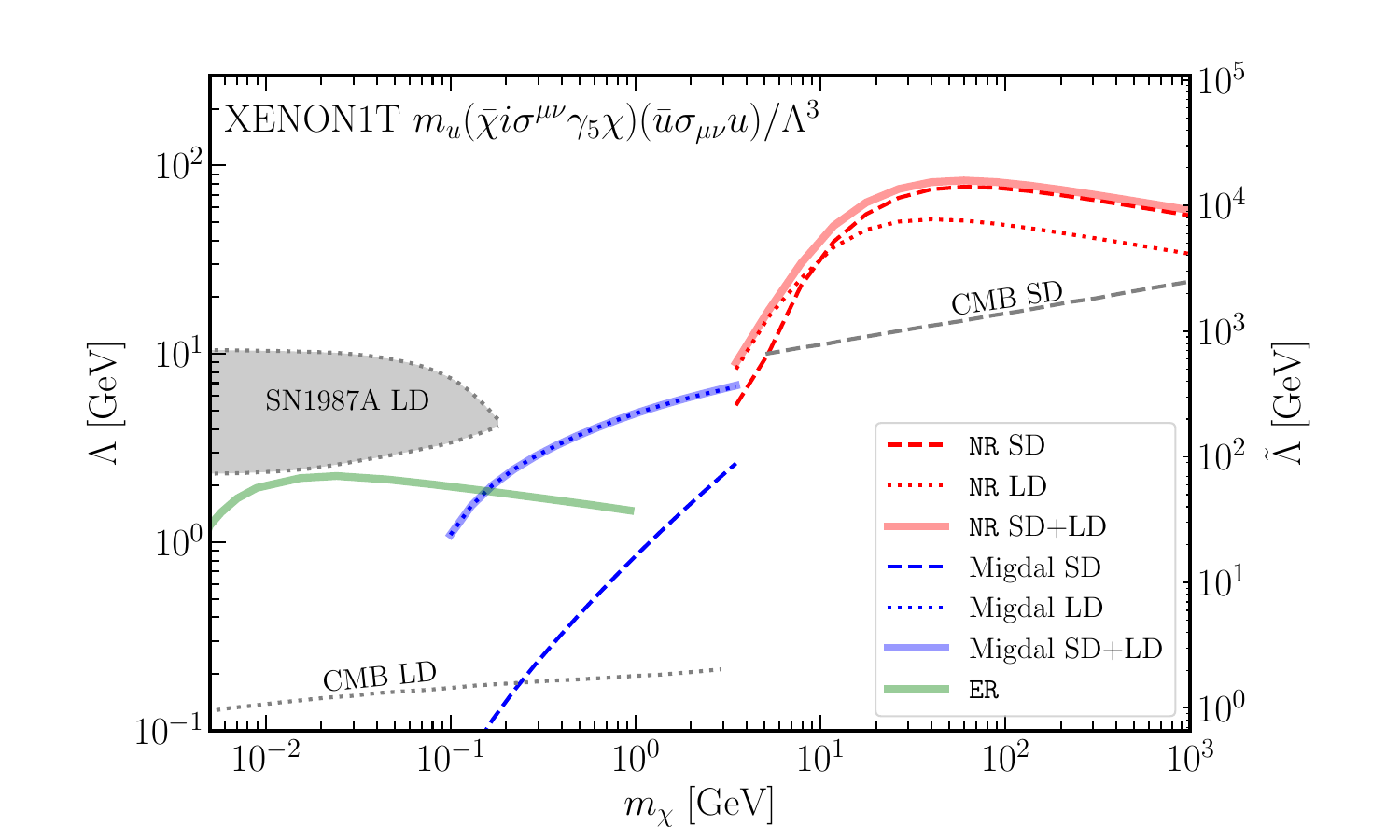}
\includegraphics[width=0.5 \textwidth]{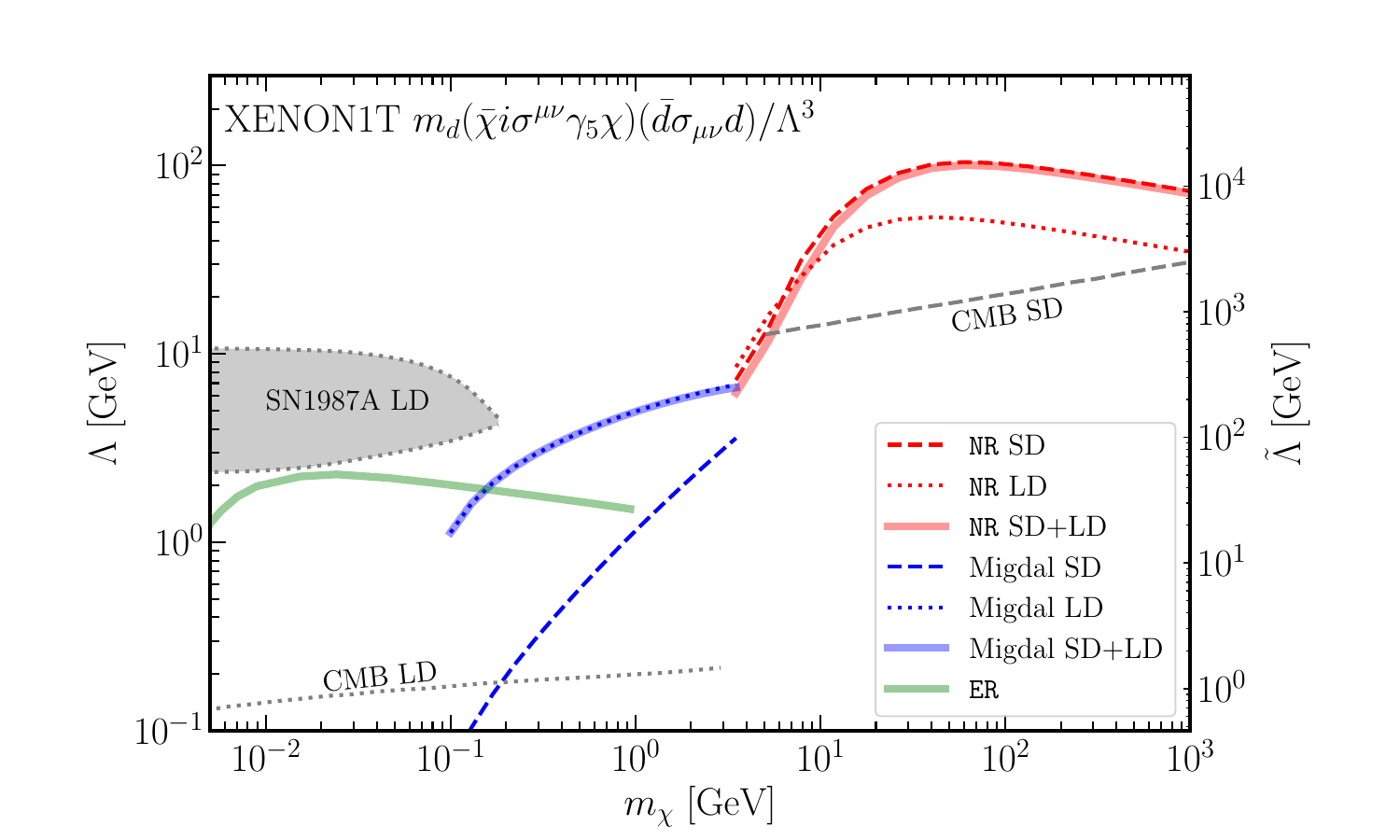}
\\
\includegraphics[width=0.5 \textwidth]{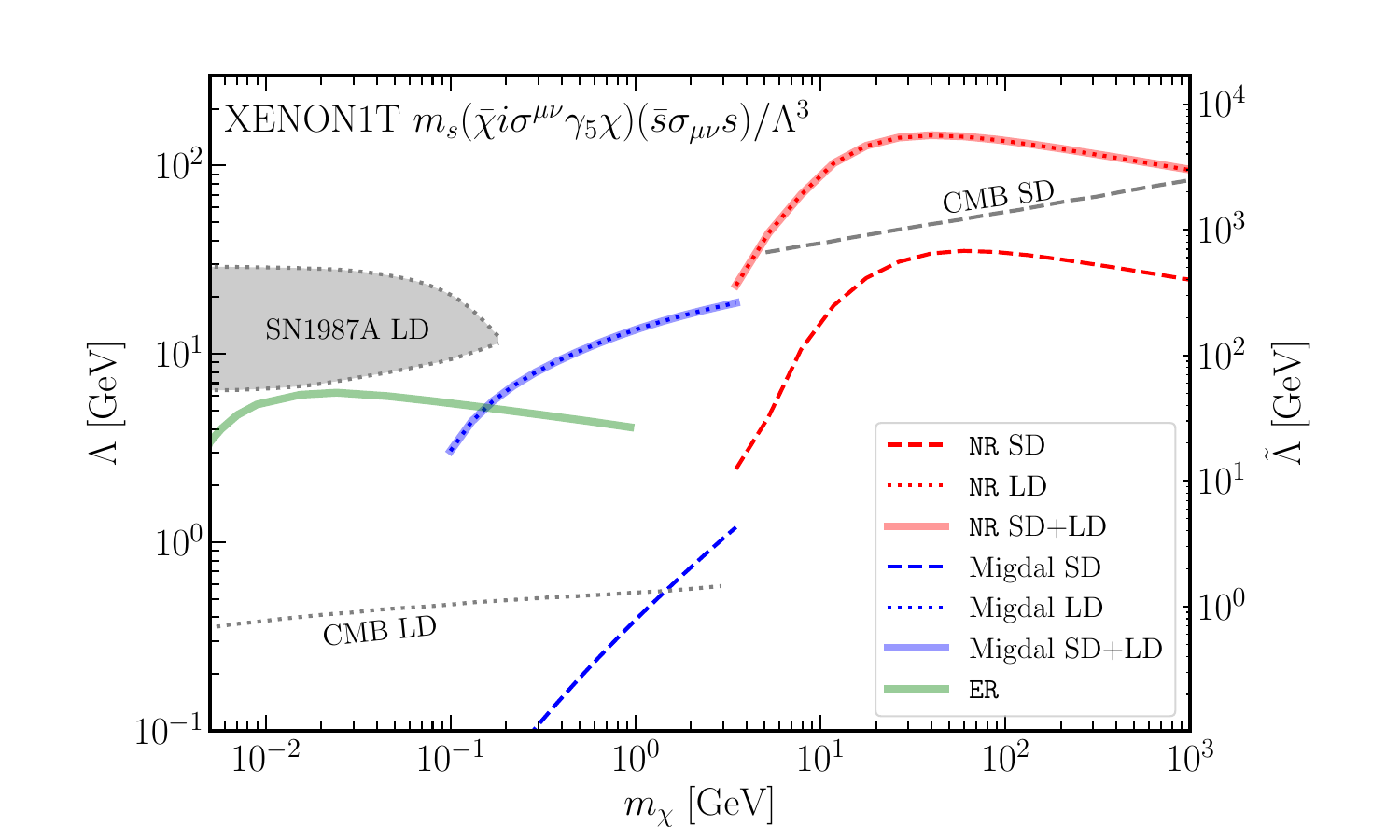}
\includegraphics[width=0.5 \textwidth]{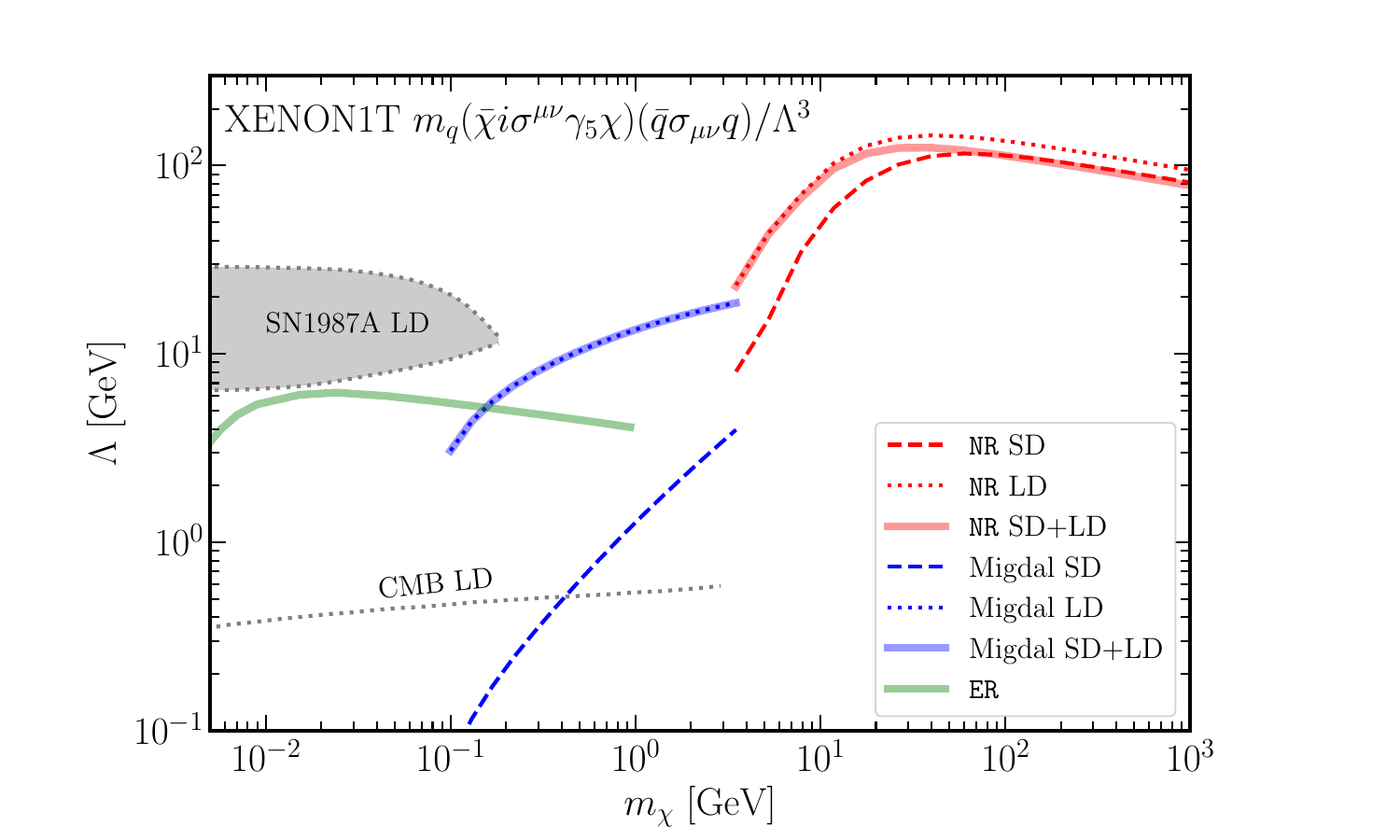}
\vspace{-5mm}
\caption{
Same as \cref{fig:Xenon1T-T1} but for the operator $\OTb$.
}
\label{fig:Xenon1T-T2}
\end{figure}

%%%%%%%%%%%%%%%%%%%%%%%%%%%%%%
\subsection{Constraint from the Migdal effect}
%%%%%%%%%%%%%%%%%%%%%%%%%%%%%%

In the DM low-mass region (i.e.\ sub-GeV region)
the constraint from $\NR$ signals loses 
sensitivity since the nucleus cannot
gain enough recoil energy 
to reach the threshold of a detector.
This dilemma can be alleviated by 
taking advantage of the Migdal effect.
In addition to the $\NR$ signals 
from DM-nucleus scattering,
the Migdal effect results in additional ionization energy $ E_{\rm EM}$ deposited in the detector
so that the total detected energy is $E_{\rm det} = \mathcal{L}  E_R + E_{\rm EM}$.
Unlike the $\ER$, a large fraction of $\NR$  energy becomes unobservable heat.
Here $\mathcal{L}$ is the quenching factor for the   $\NR$   signals, 
which accounts for the fraction of   $\NR$   energy converting into photoelectric signals.
In calculations of the Migdal effect,
it has become customary to take a constant value of $\mathcal{L}=0.15$ \cite{Bell:2021zkr}.

Taking into account the Migdal effect for the DM-nucleus scattering,
one needs to attach an additional ionization form factor to \cref{eq:NR}
to obtain the differential event rate, namely \cite{Ibe:2017yqa}
\begin{align}
\frac{dR_{\tt Migdal}}{dE_{\rm det}} 
=\frac{\rho_\chi}{m_\chi}{1\over m_A}  \int_0^{E_R^{\rm max}} d E_R 
\int_{v_{\rm min}}^{v_{\rm max}} dv F(v) v 
{d\sigma_{T}\over dE_R} (v,E_R) |Z_{\rm ion}(E_R,  E_{\rm EM})|^2,
\end{align}
where $E_R^{\rm max}= 2 \mu_{A \chi}^2 v_{\rm max}^2/m_A$ and $E_{\rm EM} = E_{\rm det}-\mathcal{L}E_R$.
Unlike the $\NR$ case, here $v_{\rm min} = (m_A E_R+ \mu_{A\chi} E_{\rm EM})/(\mu_{A\chi} \sqrt{2 m_A E_R })$
depends on not only the $\NR$ energy $E_R$ but also the ionization energy $E_{\rm EM}$.
The ionization factor $|Z_{\rm ion}|^2$ also depends on both, which is given by \cite{Ibe:2017yqa}, 
\begin{align}
|Z_{\rm ion}|^2
= \frac{1}{2 \pi} \sum_{n, \ell} \frac{d}{d E_e} p_{q_e}^c(n \ell \rightarrow E_e),
\end{align}
where $E_e$ is the kinetic energy of the ionized electron given by $E_e=E_{\rm EM} - |E_{n \ell}|$, 
with $|E_{n\ell}|$ being the binding energy of the electron labeled 
by the principal and orbital quantum numbers $n,\ell$. 
We adopt the ionization probability $p_{q_e}^c$ of the Xenon atom given in \cite{Ibe:2017yqa}.

For the Migdal effect,
we calculate the constraint with
the S2-only data set given in \cite{XENON:2019gfn,XENON:2019zpr}
for an exposure of $w=$ 22 ton-day.
The 90\% C.L. constraint is obtained via the criterion of $N_{\tt Migdal}^s < 49$ \cite{Bell:2021zkr,Tomar:2022ofh},
where the number of signal events induced by the Migdal effect is calculated by
\begin{align}
N_{\tt Migdal}^s = w \int_{0.186 \,\rm keV}^{3.90 \,\rm keV} \epsilon_{\tt Migdal}(E_{\rm det}) \frac{dR_{\tt Migdal}}{dE_{\rm det}}.
\end{align}
Here we adopt the signal efficiency of the S2-only data,
$\epsilon_{\tt Migdal} (E_{\rm det})$, given in \cite{XENON:2019gfn,XENON:2019zpr}.

The XENON1T constraints on $\OTa$ 
and $\OTb$ from the Migdal effect
are shown as blue curves in \cref{fig:Xenon1T-T1} and \cref{fig:Xenon1T-T2}.
Upon including the LD contribution,
the constraints on $\OTb$ achieve a significant enhancement, particularly for 
$\Lambda_s$, by a factor up to two orders of magnitude.
As for the constraints on $\OTa$,
there exist mass regions where the SD and LD contributions are comparable for the $u$ and $d$ quarks,
in which the SD and LD contributions have constructive interference.

%%%%%%%%%%%%%%%%%%%%%%%%%%%%%%
\subsection{Constraint from $\ER$ signals }
%%%%%%%%%%%%%%%%%%%%%%%%%%%%%%

The constraints on the edm and mdm of DM from $\ER$ signals have been carried out by the XENON1T collaboration \cite{XENON:2021qze},
in which S2-only signals induced by a single electron are used to achieve a lower energy threshold.
For the  DM-electron scattering induced by $\OTa$ and $\OTb$, there is only LD contribution. 
Hence, we can directly convert these constraints on DM dipole moments into those on $\OTa$ and $\OTb$ via \cref{eq:di-mom-Lambda}.

The XENON1T constraints on $\OTa$ and $\OTb$ from $\ER$ 
are shown as green curves in \cref{fig:Xenon1T-T1} and \cref{fig:Xenon1T-T2}.
These $\ER$ constraints effectively probe low-mass DM down to approximately 5 MeV. The small momentum transfer 
in DM-electron scattering for such low-mass DM results in an enhanced scattering cross section 
when DM interacts with electrons via edm, 
leading to a stronger constraint on $\OTb$ 
compared to $\OTa$. Moreover, the constraints on $\Lambda_s$ are more stringent than
those on $\Lambda_u$ and $\Lambda_d$
due to the quark mass factor in the definition of $\OTa$ ($\OTb$).
For the flavor conserving case, the constraints on both operators 
from Migdal effect and $\ER$ are similar
to those in the single strange quark case, 
since they are dominated by LD contributions which are further enhanced by the strange mass. 

The stronger constraint on $\OTb$ can be understood as follows. In the nonrelativistic limit,
the spin-averaged and -summed matrix element squared of the DM-electron scattering for the mdm and edm cases are 
\begin{align}
\overline{\left|\mathcal{M}_{\chi e}(q)\right|^2}_{\rm mdm} 
\simeq 16\pi \alpha \mu_\chi^2 m_\chi^2, \quad 
\overline{\left|\mathcal{M}_{\chi e}(q)\right|^2}_{\rm edm} 
\simeq 64 \pi \alpha d_\chi^2 m_\chi^2 m_e^2/q^2.
\end{align}
Here $q$ is the momentum transfer in DM-electron scattering and has a typical value of $q \simeq \alpha m_e$ in DMDD experiments.
At such a low momentum transfer scale, $\overline{\left|\mathcal{M}_{\chi e}(q)\right|^2}_{\rm edm}$ is enhanced
by a factor of $\alpha$, namely, $\overline{\left|\mathcal{M}_{\chi e}(\alpha m_e)\right|^2}_{\rm edm} \simeq 64 \pi  d_\chi^2 m_\chi^2/\alpha$,
which thus leads to a stronger constraint on the edm of DM, as compared to the mdm of DM.

In summary, due to the LD contribution from nonperturbative QCD effects of the tensor operators, the XENON1T experiment extends the sensitivity to 
MeV-scale DM. In the mass region probed by the Migdal effect 
($0.1 \, {\rm GeV} \lesssim m_\chi \lesssim 3 \, {\rm GeV}$),  
the inclusion of the LD contribution results in a sensitivity to the effective scale associated with $\OTb$ ($\OTa$) that is comparable to those well-studied operators like $\calO_{\chi q}^{\tt V}\equiv\bar{\chi}\gamma^\mu\chi \, \bar{q}\gamma_\mu q$ ($\calO_{\chi q}^{\tt A}\equiv\bar{\chi}\gamma^\mu\chi \, \bar{q}\gamma_\mu \gamma^5 q$) with $\tilde{\Lambda} \simeq \mathcal{O}(10^2) \, \rm{GeV}$ ($\tilde{\Lambda} \simeq \mathcal{O}(1) \, \rm{GeV}$) \cite{Tomar:2022ofh}. 
For the NR signal region ($5\,{\rm GeV} \lesssim m_\chi \lesssim 10^3\,\rm GeV$), the constraint on the effective scale associated with $\calO_{\chi q}^{\tt V}$ ($\tilde{\Lambda} \simeq 5 \times 10^4 \, \rm{GeV}$) is stronger than those of $\OTa$ ($\tilde{\Lambda} \simeq \mathcal{O}(10^3) \, \rm{GeV}$) and $\OTb$ ($\tilde{\Lambda} \simeq \mathcal{O}(10^4) \, \rm{GeV}$), which are further stronger than that of  $\calO_{\chi q}^{\tt A}$ ($\tilde{\Lambda} \simeq 50 \, \rm{GeV}$) \cite{Tomar:2022ofh}.

%%%%%%%%%%%%%%%%%%%%%%%%%%%%%%
\section{Constraints from other direct detection experiments }
\label{sec:full_cons}
%%%%%%%%%%%%%%%%%%%%%%%%%%%%%%

Besides the XEONON1T experiment \cite{XENON:2021qze},
constraints on DM EM dipole moments 
are also derived in XENON10 and DarkSide50 via $\ER$ signals \cite{Catena:2019gfa}
and PandaX via $\NR$ signals \cite{PandaX:2023toi}.
Hence, it is instructive to recast these constraints 
via \cref{eq:di-mom-Lambda} to fully constrain the two tensor operators, 
$\OTa$ and $\OTb$.\footnote{
Note that a consistent calculation on the constraint from PandaX $\NR$ also needs to consider the SD contribution,
as we did with XENON1T $\NR$ in the previous section. For simplicity, we restrict our discussion here to the LD contribution only.
}
Together with the constraints from XENON1T calculated above, 
we show in \cref{fig:compare} all the available bounds on $\OTa$ and $\OTb$ obtained
in this work by colored curves. 
Here we only show the results for the flavor conserving case 
to facilitate comparison with the results in the literature 
(represented by dashed gray curves for the $\NR$ \cite{Kang:2018rad} 
and the Migdal effect \cite{Tomar:2022ofh}), 
in which only the SD contribution is considered.\footnote{
To cross check our calculations, we attempted to reproduce the XENON1T constraints in \cite{Tomar:2022ofh}
on the two DM-quark tensor operators from the Migdal effect. We found that we could get results consistent 
with \cite{Tomar:2022ofh} only when we adopted the mistaken value of $g_T^{s/N}= -0.027$,
as discussed in footnote\,\ref{foot:gTs}.
}

\begin{figure}%[htbp]
\includegraphics[width=0.5 \textwidth]{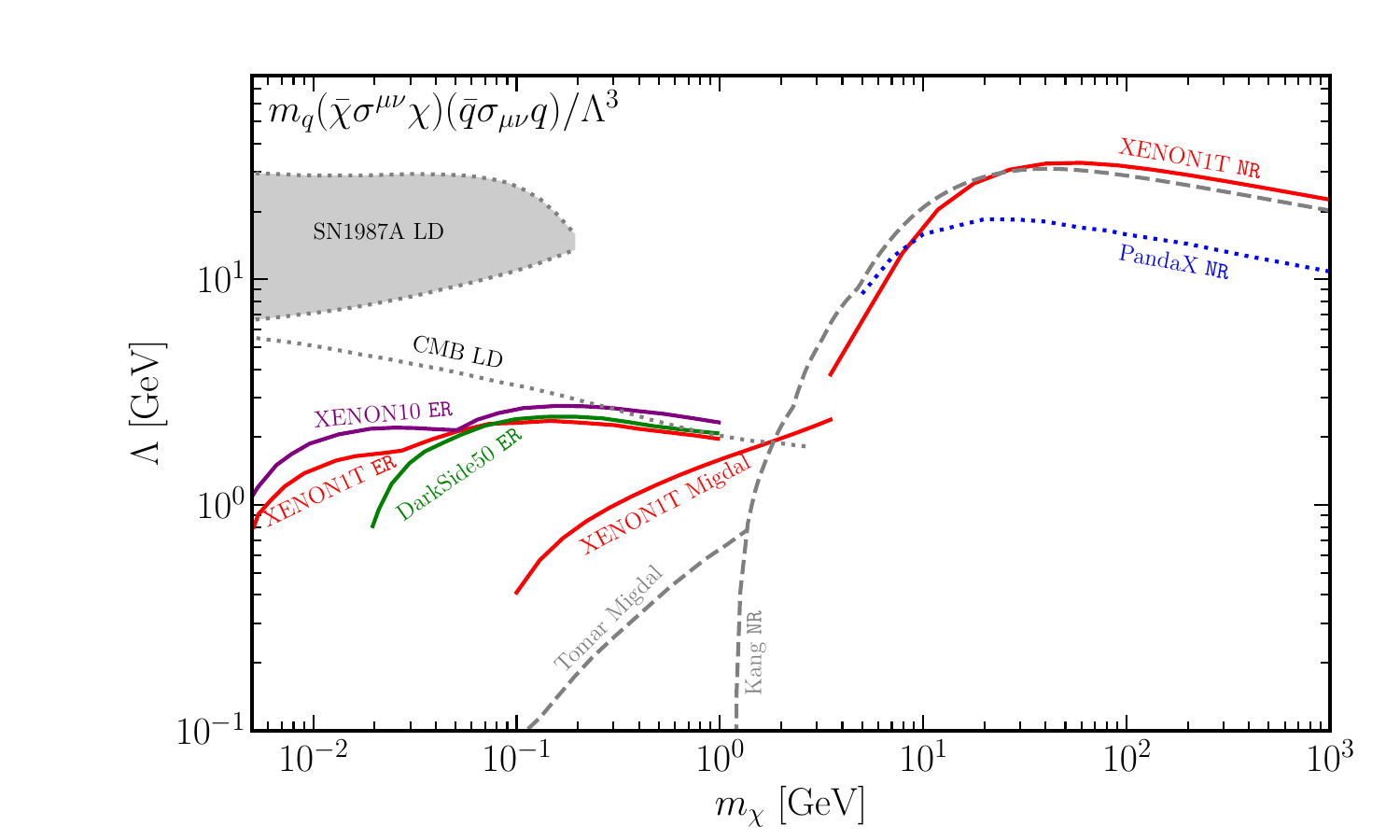}
\includegraphics[width=0.5 \textwidth]{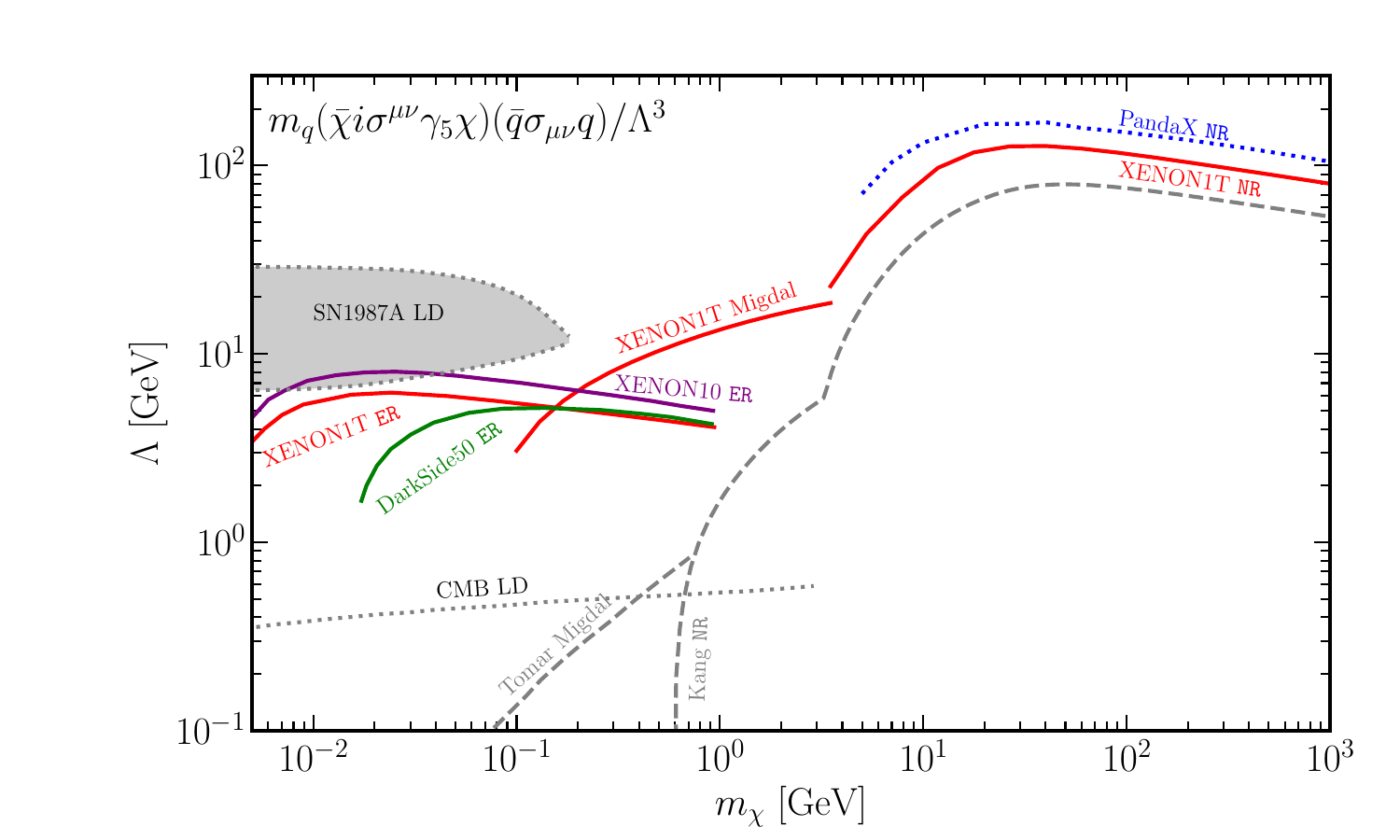}
\vspace{-5mm}
\caption{Comparison of constraints on $\OTa$ ({\it left panel}) and $\OTb$ ({\it right panel}) for the flavor conserving case from current DMDD experiments. Colored lines show new constraints in this work with the LD contribution included. Two gray dashed curves show previous constraints from the $\NR$ \cite{Kang:2018rad} and the Migdal effect \cite{Tomar:2022ofh}, in which only the SD contribution is considered.
Two gray dotted lines show constraints from SN1987A and CMB discussed in \cref{sec:nonDD}, in which only the LD contribution is considered. The legends follow those in \cref{fig:Xenon1T-T1} and \cref{fig:Xenon1T-T2}: the dashed, dotted, and solid curves indicate the constraints with SD-only, LD-only, and full contributions respectively.}
\label{fig:compare}
\end{figure}

By taking advantage of the new LD contribution, $\OTa$ and $\OTb$ can also be constrained by the $\ER$,
in addition to the $\NR$ and the Migdal effect. These new $\ER$ constraints cover a previously uncovered
low-mass region with ${5\,\rm MeV} \lesssim m_\chi \lesssim {100\,\rm MeV}$
and surpass the previous constraints from the Migdal effect in the mass region,
${100\,\rm MeV} \lesssim m_\chi \lesssim {1\,\rm GeV}$. Owing to the enhancement of the LD contribution
at the small momentum transfer in the Migdal effect, the newly calculated XENON1T constraint
from the Migdal effect is stronger than the previous one from the Migdal effect
by about a factor of three (one order of magnitude) in the $\OTa$ ($\OTb$) case.
In particular, for the $\OTb$ case, the XENON1T constraint from the Migdal effect can be even stronger
than the previous ones from the $\NR$ in the mass region of ${0.7 \,\rm GeV}\lesssim m_\chi \lesssim {3\,\rm GeV}$.
Due to the relatively large momentum transfer in the $\NR$, the improvements in the constraints from the $\NR$ are not so obvious,
compared to those from the Migdal effect. For the $\OTb$ case,
the PandaX constraint from the $\NR$ improves in the large mass region ($m_\chi \gtrsim 3\,\rm GeV$).
But it is expected to become slightly weaker for heavier DM when the SD contribution
is also considered because of the destructive interference between the LD and SD contributions as shown in \cref{fig:Xenon1T-T2}.

%%%%%%%%%%%%%%%%%%%%%%%%%%%%%%
\section{Constraints from non-direct-detection experiments and an example of UV completion}
\label{sec:nonDD}
%%%%%%%%%%%%%%%%%%%%%%%%%%%%%%

Since the particle properties of dark matter are completely unknown, 
it is important to explore it in various types of observations and 
experiments to obtain hopefully complementary information. 
In this section we briefly discuss the constraints coming from collider searches and the supernova (SN1987A) and cosmic microwave background (CMB) observations. 

The effective operators $\OTa$ and $\OTb$ can be probed at the LHC via the mono-jet search,
in which DM produced via the process $q \bar{q} \to 
\chi \bar{\chi} +j$ appears as missing energy at collider detectors. Due to the nature of their four-fermion interactions, the signal cross section is proportional to the center-of-mass energy squared at the parton level,  
and thus gets enhanced at the LHC energy. 
For the DM mass below hundreds of GeV, one generally obtains a bound $\tilde{\Lambda}\gtrsim 1\,\rm TeV$ \cite{Belyaev:2018pqr}. 
Nevertheless, it is delicate to compare directly the bound with those extracted from DMDD experiments. First, the bound at high energy colliders is less sensitive to the DM mass as long as the latter is not too close to the parton energy. And it is insensitive to the Lorentz structure of effective interactions. A well-known example is the vastly different bounds on the spin-independent and -dependent interactions extracted in direct detection experiments which yield similar signals at colliders. Furthermore, the search at colliders cannot distinguish a tensor interaction from other structures. In contrast, direct detection at low energy allows to examine the tensor structure in a comprehensive manner: only for a tensor structure can a LD interaction be induced from a four-fermion DM-quark interaction, which results in interesting interference between the two and correlates the signal channel in $\NR$ and Migdal effect on one side and the one in $\ER$ on the other. 
The ratio of signal strengths between the two would help distinguish the tensor DM-quark interactions from other types of interactions and determine the DM mass, if a DM signal were observed. Second, the bounds set at colliders would be modified significantly by a mediator of intermediate mass between $\calO (100 ~\rm MeV)$ and $\calO (1~\rm TeV)$, which is common in various portal mechanisms~\cite{An:2012va}. The constraints from low-energy detection are not flawed with this issue for low-mass DM, and thus more robust. 

We next examine the constraints from SN1987A and CMB. Light DM particles with mass $m_\chi \lesssim 400~\rm MeV$ could be generated in pairs within the supernova core and then would escape, this would increase the supernova cooling rate and consequently affect the observed supernova neutrino spectrum. Since the generated DM could be reabsorbed by the supernova if its interaction with SM particles is sizable enough, the constraints are two-sided, resulting in an allowed region in parameter space. Regarding the CMB constraint, the extra energy injection from DM annihilation can affect the recombination history leading to modifications in the temperature and polarization power spectra of the CMB. We have converted the constraints on the DM edm and mdm in~\cite{Chu:2018qrm,Chu:2019rok} obtained from the SN1987A and CMB to those on $\OTa$ and $\OTb$ by employing \cref{eq:di-mom-Lambda}. 
In addition, the SD $\OTa$ and $\OTb$ interactions can also be directly constrained by the SN1987A and CMB. While the supernova constraint with only SD interactions has not been worked out in the literature, the case for CMB has been done in~\cite{Belyaev:2018pqr}.
Note that the CMB constraint becomes significantly weaker for $m_\chi \lesssim m_\pi$ if we include only SD interactions, as there would be no annihilation channel available at the tree level. 
All of the above constraints from the SN1987A and CMB are also included in \cref{fig:Xenon1T-T1}, \cref{fig:Xenon1T-T2}, and \cref{fig:compare} as gray regions (SN1987A LD), gray dotted lines (CMB LD), and gray dashed lines (CMB SD) respectively.\footnote{
The CMB SD curves (gray dashed lines) in \cref{fig:Xenon1T-T1} and \cref{fig:Xenon1T-T2}  represent the CMB constraints with only the SD interaction and are taken from \cite{Belyaev:2018pqr}. In \cite{Belyaev:2018pqr}, 
the constraints were obtained by parametrizing 
the two tensor-type interactions as dimension-six operators and taking 
a universal Wilson coefficient for the three light quarks. 
Since these constraints cannot be trivially translated into those for the individual quark cases,
we just simply take the constraint from \cite{Belyaev:2018pqr} as a rough comparison. Note that the actual CMB SD constraints for the individual quark cases would be somewhat weaker due to reduced contributing channels from single quark flavor. 
}
In general, the parameter region probed by the SN1987A is more strongly constrained 
but it does not overlap with those probed by DMDD experiments and the CMB. Among the latter two, the DMDD constraint is generally stronger (weaker) for the $\OTb$ ($\OTa$) interaction.

%%%%%
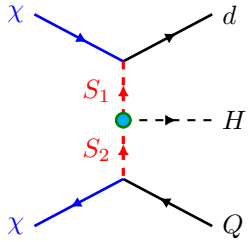
\begin{figure}[t]
\centering
\begin{tikzpicture}[mystyle,scale=1.3]
\begin{scope}[shift={(1,1)}]3
\draw[f,blue] (-1.5,1.5) node[left]{$\chi$}  -- (0,0.7);
\draw[f] (0, 0.7) -- (1.5,1.5) node[right]{$d$};
\draw[sb, red, very thick] (0,0.7) -- (0,-0.3) node[midway,xshift = -10 pt]{$S_1$};
\draw[sb, red, very thick] (0,-0.3) -- (0,-1.3) node[midway,xshift = -10 pt]{$S_2$};
\draw[s, thick] (0,-0.3) -- (1.5,-0.3) node[right]{$H$};
\draw[f,blue] (0,-1.3)--(-1.5,-2.1) node[left]{$\chi$};
\draw[f]  (1.5,-2.1) node[right]{$Q$}--(0, -1.3);
\draw[draw=Green,fill=cyan] (0,-0.3) circle (0.12cm);
\end{scope}
\end{tikzpicture}
\caption{
Feynman diagram in the $\mathbb{Z}_3$ model that induces a tensor effective interaction.
}
\label{fig:Feyn-dia2}
\end{figure}
%%%%%

Finally, we show a simple UV model that induces at the tree level a tensor-type interaction without being accompanied by a quark mass. Consider a $\mathbb{Z}_3$ DM model by extending the SM with a vector-like fermion DM $\chi(1,1,0)$, and two colored scalars $S_1(3,1,-1/3)$ and $S_2(3,2,1/6)$, where the numbers in parentheses denote the quantum numbers in the SM gauge group. Under $\mathbb{Z}_3$ symmetry, $\chi\to e^{i 2\pi/3} \chi$ and $S_{1,2} \to e^{- i 2\pi/3}S_{1,2}$, while the SM fields are intact. The relevant Lagrangian terms are 
   \begin{align}
    \calL \supset y_d (\bar d \, \chi_L) S_1 + y_Q (\bar Q \, \chi_R) S_2 + \mu_S H^\dagger S_1^\dagger S_2 + {\text{h.c.}},
\end{align}
where $y_d$ and $y_Q$ are the dimensionless Yukawa couplings, $\mu_S$ is the dimensionful triple coupling for the scalars, and $H$ is the SM Higgs field with vacuum expectation value $v$. The Feynman diagram shown in~\cref{fig:Feyn-dia2} results in the effective interaction upon integrating out heavy scalars $S_1$ and $S_2$:
\begin{align}
\nonumber
& 
{{y_d \, y_Q^* \, \mu_S \, v} \over {\sqrt{2} m_{S_1}^2} m_{S_2}^2} (\overline{d_R}\chi_L) (\overline{\chi_R} d_
L)
\\
= 
& -{1\over 2}{{y_d \, y_Q^* \, \mu_S \, v} \over {\sqrt{2} m_{S_1}^2} m_{S_2}^2} 
\left[ (\overline{d_R} d_L) (\overline{\chi_R} \chi_
L) + {1\over 4} (\overline{d_R} \sigma_{\mu\nu} d_L) (\overline{\chi_R} \sigma^{\mu\nu} \chi_
L) \right],
\end{align}
where the Fierz identity is applied. It is natural to assume $m_{S_1}\simeq m_{S_2}\simeq\mu_S=M$ so that the induced tensor operator has a Wilson coefficient $y_d y_Q^* v/(8\sqrt{2} M^3)$ whose magnitude is defined as $\tilde\Lambda^{-2}$ in our discussion. 

%%%%%%%%%%%%%%%%%%%%%%%%%%%%%%
\section{Conclusion}
\label{sec:conc}
%%%%%%%%%%%%%%%%%%%%%%%%%%%%%%

In this work, we offered a more complete investigation of the two DM-quark tensor operators
in dark matter direct detection (DMDD) experiments in the framework of chiral perturbation theory.
We found that DM-quark tensor operators can induce electromagnetic dipole moment operators of the DM,
in addition to the well-studied DM-nucleon 4-fermion operators. In previous calculations for DMDD experiments,
the constraints on DM-quark tensor operators were obtained by calculating the $\NR$
and the Migdal effect induced by these 4-fermion operators. The DM dipole moment operators give rise to new contributions for both DM-electron and DM-nucleus scatterings. Consequently,
a consistent calculation of the constraints from the $\NR$ and Migdal effect should include
both DM-nucleon and DM dipole moment operators, which may result in interesting interference effects.
Remarkably, the DM-quark tensor operators can also be constrained by $\ER$ signals caused
by the newly studied DM-electron scattering. 
In this manner, we derived the constraints from the $\NR$ and the Migdal effect using the XENON1T data,
and recast the existing bounds from the $\ER$ (XENON10, XENON1T, DarkSide50) and $\NR$ (PandaX) signals
to yield comprehensive constraints on the tensor interactions.
Our results have improved significantly over the previous ones in the literature, especially in the sub-GeV region.

%%%%%%%%%%%%%%%%%%%%%%%
\section*{Acknowledgement}
\addcontentsline{toc}{section}{\numberline{}Acknowledgements}
%%%%%%%%%%%%%%%%%%%%%%%

This work was supported in part by the Guangdong Major Project of Basic
and Applied Basic Research No.\,2020B0301030008, and by the Grants 
No.\,NSFC-12035008, 
No.\,NSFC-12247151, 
No.\,NSFC-12305110,
and No.\,NSFC-12347121.

%%%%%%%%%%%%%%%%%%%%%
%\newpage
\appendix
%%%%%%%%%%%%%%%%%%%%%

%%%%%%%%%%%%%%%%%%%%%%%%%%%%%%
\section{Squared form factor}
\label{app:sform}
%%%%%%%%%%%%%%%%%%%%%%%%%%%%%%

The squared form factors $F_{i,j} ^{(N,N^\prime)}(\pmb{q}^2,\pmb{v}^{\perp 2}_{T})$ defined in \cref{eq:samp}
are related to the basic independent nuclear form factors in the following way \cite{Fitzpatrick:2012ix},
\begin{align}
F_{1,1}^{(N,N^\prime)} & =  
F_{M}^{(N,N^\prime)},
\\
F_{4,4}^{(N,N^\prime)} & = 
\frac{1}{16}\left( F_{\Sigma^\prime}^{(N,N^\prime)}
+ F_{\Sigma^{\prime\prime}}^{(N,N^\prime)} \right),
\\
F_{5,5}^{(N,N^\prime)} & = 
\frac{\pmb{q}^2}{4}\left({\pmb{v}_T^\perp}^2 F_{M}^{(N,N^\prime)}
+ \frac{\pmb{q}^2}{m_N^2} F_{\Delta}^{(N,N^\prime)} \right),
\\
F_{6,6}^{(N,N^\prime)} & = 
\frac{\pmb{q}^4}{16} F_{\Sigma^{\prime\prime}}^{(N,N^\prime)},
\\
F_{4,5}^{(N,N^\prime)} & = 
- \frac{\pmb{q}^2}{8 m_N} F_{\Sigma^{\prime},\Delta}^{(N,N^\prime)},
\\
F_{4,6}^{(N,N^\prime)} & =
\frac{\pmb{q}^2}{16} F_{\Sigma^{\prime\prime}}^{(N,N^\prime)},
\\
F^{(N,N^\prime)}_{10,10} & =
\frac{\pmb{q}^2}{4} F_{\Sigma^{\prime\prime}}^{(N,N^\prime)},
\\
F^{(N,N^\prime)}_{11,11} & =
\frac{\pmb{q}^2}{4} F_{M}^{(N,N^\prime)},
\\
F^{(N,N^\prime)}_{12,12} & =  
\frac{{\pmb{v}_T^\perp}^2}{16}\left(\frac{1}{2}F_{\Sigma^{\prime}}^{(N,N^\prime)}
+ F_{\Sigma^{\prime\prime}}^{(N,N^\prime)}\right)
+ \frac{\pmb{q}^2}{16 m_N^2}\left( F_{\Tilde{\Phi}^{\prime}}^{(N,N^\prime)}
+ F_{\Phi^{\prime\prime}}^{(N,N^\prime)} \right),   
\\
F^{(N,N^\prime)}_{11,12} & = 
-\frac{\pmb{q}^2}{8 m_N} F_{M,\Phi^{\prime\prime}}^{(N,N^\prime)}.
\end{align}

%%%%%%%%%%%%%%%%%%%%%%%%%%%%%%
%%%%%%%%%%%%%%%%%%%%%%%%%%%%%%
\bibliography{refpaper.bib}{}
\bibliographystyle{utphys28mod}

\end{document}